\documentclass[12pt,onecolumn]{IEEETran}
\makeatletter
\def\endthebibliography{%
	\def\@noitemerr{\@latex@warning{Empty `thebibliography' environment}}%
	\endlist
}
\makeatother

\usepackage{amsmath}
\usepackage{amsthm}
\usepackage{amsfonts}
\usepackage{epsfig}
\usepackage{amssymb}
\usepackage{color}
\usepackage{cite}
\usepackage{subfigure}
\usepackage{multirow}
\usepackage{rotating}
\usepackage{graphicx}
\usepackage{tabularx}
\usepackage{array}
\usepackage{url}
\usepackage{epstopdf}
\usepackage{theoremref}
\usepackage{graphicx}
\usepackage{subfigure}
\usepackage{algorithm} 
\usepackage{algorithmic}
\usepackage[algo2e]{algorithm2e} 
\usepackage{setspace}
\usepackage{float}

\DeclareMathOperator*{\argmax}{arg\,max} 
\newcolumntype{C}[1]{>{\centering\let\newline\\\arraybackslash\hspace{0pt}}m{#1}}

\definecolor{mypink1}{cmyk}{0, 0.7808, 0.4429, 0.1412}
\date{}

\begin{document}
	
	\title{ \LARGE  AI-Based  Secure NOMA and Cognitive Radio enabled Green Communications: Channel State Information and Battery Value Uncertainties  }
	\author{Saeed~Sheikhzadeh, Mohsen~Pourghasemian, Mohammad~R.~Javan, Nader~Mokari	and Eduard A. Jorswieck
		\footnote{Saeed sheikhzadeh, Mohsen Pourgasemian, and  Nader Mokari are with Department of Electrical and Computer Engineering, Tarbiat Modares University, Tehran, Iran (e-mail: nader.mokari@modares.ac.ir). Mohammad~R.~Javan is with Department of Electrical and Computer Engineering, Shahrood University of Technology, Iran (e-mail: javan@shahroodut.ac.ir). Eduard A. Jorswieck is with the Department of Information Theory and Communication Systems, Technische Universit\" at Braunschweig, Braunschweig, Germany (e-mail: jorswieck@ifn.ing.tu-bs.de). This work was
			supported by the joint Iran national science foundation (INSF) and German research foundation (DFG) under grant No. 96007867.
		}	 
	}
	
	\maketitle

\begin{abstract}
In this paper, the security-aware robust resource allocation in energy harvesting cognitive radio networks is considered with cooperation between two transmitters while there are uncertainties in channel gains and battery energy value. 
To be specific, the primary access point harvests energy from the green resource and uses time switching protocol to send the energy and data towards the secondary access point (SAP). Using power-domain non-orthogonal multiple access technique, the SAP helps the primary network to improve the security of data transmission by using the frequency band of the primary network. 
In this regard, we introduce the problem of maximizing the proportional-fair energy efficiency (PFEE) considering uncertainty in the channel gains and battery energy value subject to the practical constraints. Moreover, the channel gain of the eavesdropper is assumed to be unknown.  Employing the decentralized partially observable Markov decision process, we investigate the solution of the corresponding resource allocation problem. We exploit multi-agent with single reward deep deterministic policy gradient (MASRDDPG) and recurrent deterministic policy gradient (RDPG) methods. These methods are compared with the state-of-the-art ones like multi-agent and single-agent DDPG. Simulation results show that both MASRDDPG and RDPG methods, outperform the state-of-the-art methods by providing more PFEE to the network.\\
%It is worth mentioning that  %When there is no uncertainty, the RDGP and MASRDDPG achieve nearly the same performance compared with two other methods. Moreover, in the existence of 
%with 10\% uncertainty, the RDPG performs nearly 14\% better than MASRDDPG.

\end{abstract}

	\section{Introduction}\label{Introduction}

	\subsection{{Background}}

	Cognitive radio (CR) is a promising way to improve the spectrum efficiency due to the ability of the system to reassign  unoccupied frequencies \cite{Liang1}. In the CR network, there are two kinds of network nodes, a primary that is known as the owner of the spectrum and a secondary which uses the frequency based on different policies. These policies can be divided into three categories as interweave, underlay, and overlay \cite{Xu3}. In the interweave and underlay CR network, the secondary is given permission to use the spectrum when there is vacancy in frequency band and an acceptable interference to the primary receiver, respectively. Moreover, in the overlay case, the secondary network cooperates with the primary in order to improve the primary throughput. The overlay case is also called as cooperative cognitive which is widely studied in \cite{Urgaonkar1,Xu4}.  
	\\
	Wireless powered communication networks (WPCNs) are considered often which are promising techniques to provide a stable energy resource in energy limited devices \cite{Bi1, Chingoska1}. Recently, this technique is combined with CR networks as cognitive wireless powered communication networks (CWPCNs) \cite{Xu3, Zheng1, He1}. There are two popular energy harvesting (EH) techniques which are used in CWPCNs, time-switching (TS) and power-splitting (PS) schemes \cite{Ni1, Xu2}. 
	\\
	In order to prolong the network lifespan in energy-limited devices and decrease the harmful effect of energy production  on the environment, energy efficient (EE) resource allocation is introduced in \cite{Song1,Li2}. In addition, the combination of the EE resource allocation problem and the CWPCNs is adopted in \cite{Zhou2,Zhang5} which improves the energy consumption at power restrictive cognitive transmitters. The popular EE resource allocation problems can be divided into four main categories such as global EE, proportional fair EE (PFEE), max-min fair EE (MMFEE), and harmonic fair EE (HFEE) methods \cite{Guo1}. In the global EE, the ratio of rate to energy is maximized. Moreover, in  PFEE, MMFEE, and HFEE, the objectives are maximizing the logarithm of EE, maximizing the minimum value of EE, and minimizing the reverse of EE, respectively. 
	\\
	The uncertainty in the telecommunication system is inevitable. The assumption of perfect information about environment or the battery value in devices due to the stochastic nature of wireless channel, delays or failure of circuit of devices mostly is impossible. The unavailability or imperfection of these kinds of information leads to performance degradation of the network and subsequently, the network efficiency decreases. The uncertainty may exist in any section of the communication network like channel gain \cite{Hasan1,Hasan2}, locations of devices \cite{Muppirisetty1}, and demand and capacity of available resources \cite{Johansson1}. 
	\\
	Moreover, power domain non-orthogonal multiple access (PD-NOMA) is one of the main candidates for improving the next generation of cellular network. PD-NOMA can easily enhance throughput due to high spectral efficiency (SE). In contrast to orthogonal multiple access (OMA) that only assigns one resource block to each user, the main idea of PD-NOMA is to exploit one orthogonal channel resource block for more than one user to improve SE \cite{Ding1,Ding2}.

	\subsection{Literature Review}
	%%%%%%%%%%%%%%%%%%%
	Recently, there have been many works which address the challenges of using the EH in wireless communications. We categorize the the aforementioned studies in the following groups:   
	
	%The combination of EH in downlink and uplink transmission by using NOMA and full-duplex techniques in the uplink are investigated in \cite{Do1}. 
	%EH has been considered in several papers. We categorize the related works into two groups as follows:
	
	\begin{itemize} 
		\item \textbf{EH and EE}: The joint EH and EE maximization problem is  investigated in \cite{Kuang1,Zhang3,Gupta1,Xu2,Zhou2}. 
		The authors in \cite{Kuang1} try to maximize EE of D2D link by considering the QoS of users and the causality constraint of D2D links. %\cite{Zhou2} proposes the harvest then transmit protocol for machine to machine (M2M) communication network. By considering the EE, they jointly optimize channel selection, peer discovery, power control, and time scheduling subject to the causality and QoS constraints.
		 \cite{Zhang3} proposes an EE maximization problem in heterogeneous networks based on simultaneous wireless information and power transfer (SWIPT) and NOMA. In this paper, the power allocation and sub-channel matching resource allocation problem subject to the interference threshold is investigated. Moreover, the devices convert the received signal of other BSs into energy. \cite{Gupta1} investigates EE maximization problem in SWIPT AF relay networks. The authors formulate their proposed problem with the goal of optimizing the source and relays total transmit powers, joint sub-carrier and user assignment, and relay selection and TS coefficient determination subject to total transmit power budget of source and energy casualty constraints of relays. The combination of robust EE and EH in heterogeneous networks where each femtocell adopts SWIPT is considered in \cite{Xu2}. The authors propose the EE maximization problem subject to outage probability of macrocell users (MUs) to optimize transmit power allocation and PS factor.
		\item \textbf{Secrecy and EH:} \cite{Hu1,Wang4,Xu3,Zhao3} consider the combination of secrecy and EH. The physical security in SWIPT AF relay network is considered in \cite{Hu1}, while the imperfect channel gain between relays and the eavesdroppers is modeled by random CSI errors. The objective is secure data transmission with constraints on relays transmit powers and the eavesdropper signal to interference plus noise ratio (SINR). Moreover, there is no direct link between the transmitter and the receiver, and the eavesdropper can only listen to the signals that are transmitted by relays. The authors of \cite{Wang4} consider NOMA technique and nonlinear EH model to maximize the secrecy energy efficiency. In \cite{Xu3}, the CR network is considered to maximize the secondary users (SUs) ergodic rate. Furthermore, the SUs,  which are not supposed to transmit data, are considered to transmit artificial noise toward the eavesdroppers.
		\item \textbf{Artificial Intelligence and EH}: Artificial Intelligence (AI)-based EH problem with a cooperating node between source-destination pairs is proposed in \cite{Zhang4}. In this work the relay decides which link at each time slot has permission to transmit under the EH constraint. The authors of \cite{Qiu1} consider the effective energy management in wireless sensor network by utilizing the Deep Deterministic Policy Gradient (DDPG) reinforcement learning method. In \cite{Li4}, the problem of long term throughput maximization in point to point communication is solved by using the DDPG method.
		\item \textbf{AI and uncertainty}: \cite{Yang2} studies the effect of uncertainty in channel gains in wireless secure communication system with coexistence of intelligent reflecting surface (IRS). The authors propose the post decision state (PDS) and prioritized experience replay (PER) techniques to improve the learning efficiency as well as performance of secure communication. However, they consider that the BSs know CSI of the eavesdropper which is not applicable in the real environment.
	\end{itemize}
	
Various optimization problems with different technologies and restrictions have been widely studied in the EH secure wireless communication network. Most of these works use an iterative algorithm to find the optimal or sub-optimal solutions. These solutions mostly involve high computational complexity which can not be deployed in sensors or small nodes. Moreover, due to the dynamic behavior of wireless channel conditions, these algorithms have to be updated in a very short time duration which is impractical in real environment. 
	
The problems of uncertainty in the channel gain and other parameters of the network are still the bottleneck in wireless communication. This situation gets worse when an eavesdropper with unknown channel gain enters into the system. By considering these issues, machine learning techniques can be considered as a great way to learn the condition of the network and solve the problem in a timely manner. The only recent paper which uses machine learning algorithm to deal with the channel  gain uncertainty in cognitive radio network is  \cite{Yang2}. The authors adopt single agent reinforcement learning method in which the state of the environment is fully observable to the agent, even the eavesdropper channel gain, to decide the appropriate action. On the other hand, considering the uncertainty constraints as probabilistic expressions provides more robustness to the solution which is neglected in their formulation.
	\\
	Motivated by these considerations, in this paper we formulate the EH-enabled CR network with Time Switching PD-NOMA (TS-NOMA) technique, as a multi agent partially observable Markov decision process (POMDP) in an uncertain environment, considering the existence of an eavesdropper. The partially observability means that the agent does not have full state information of the channel gains of some devices, i.e., the eavesdropper, and the uncertainty means  the agent knows the battery energy level and channel gains with bounded errors. We consider three different formulations for the uncertainty in our environment as the worst case, probabilistic, and Bernstein approximation, and formulate our problem based on these three schemes. Then we utilize two deep reinforcement learning methods which are suitable for partially observable and uncertain environment as 1) multi agent deep deterministic policy gradient with single reward (MASRDDPG) and 2) deep recurrent deterministic policy gradient (RDPG) as a solution for our problem.

	\subsection{Contribution}
	
	 To the best of the authors' knowledge, considering the CR EH-enabled with uncertain conditions and partially observable environment to overcome the uncertainty of the information has not been done by previous AI-based researches.
	Accordingly, the main contributions of this paper are summarized as follows: 
	\begin{itemize}
		\item We investigate PD-NOMA cooperative cognitive TS based EH system with the existence of an eavesdropper, and the uncertainty of channel gains, the amount of harvesting energy, and the battery energy level. To this end, we consider that there is one primary access point (PAP) that sends information to the primary user equipment (PUE). However, due to poor channel condition, the secondary access point (SAP) which harvests energy by the TS  techniques, helps to forward data of the PUE to achieve opportunity of transmitting its own information. The eavesdropper tries to listen to the transmitted data in the network.
		\item We formulate EE optimization problem in which the objective is to maximize the secure PFEE by finding transmit powers and the TS factor in  the partially observable environment
			with uncertain values, while considering battery energy constraints, the QoS requirement of the PUE, energy causality limitations, and the total transmit power budget.
		\item 
		By considering uncertainties in the channel gains and the battery values, we adopt the worst case, the probabilistic, and the Bernstein approximation approaches for the uncertainties and reformulate the resource allocation problem. Then we utilize RDPG method, which is suitable for the partially observable and uncertain environment. In addition, by considering different objectives for the PAP and the SAP along with their common objective to serving the PUE, we adopt the MASRDDPG method to cope with the single and shared objectives besides the partially observability.
		\item  By conducting numerical simulations, when considering the single reward along with shared reward for multi agent method, the MASRDDPG outperforms by 5.6\%, 22.7\%, and 31.8\% compared to RDPG, MADDPG, and DDPG methods, respectively, from the average secrecy rate point of view. On the other hand, by the expense of time complexity, using recurrent neural network (RNN) to support the uncertainty and partially observability in the environment, the RDPG method outperforms the MADDPG and DDPG by 16.4\% and 25.2\%, respectively, from the average secrecy rate point of view.
	\end{itemize}

	\subsection{Structure}
	The rest of this paper is organized as follows. 
	In Section \ref{system_model}, we present our proposed the TS-NOMA system model with cooperative transmission and the EH capability. %In Section \ref{Background_Machine_Learning}, the machine learning methods which are used in this paper are investigated.
	 In Section \ref{Solution_algorithm}, we propose a navel solution and algorithm to find an optimal energy consumption of network. In Section \ref{Simulation_result}, the simulation results are provided. Finally, Section \ref{conclusion} concludes the paper.

	\section{System model and problem formulation} \label{system_model}
	
	\begin{figure}[tb]
		\begin{center}
			\includegraphics[width=14 cm]{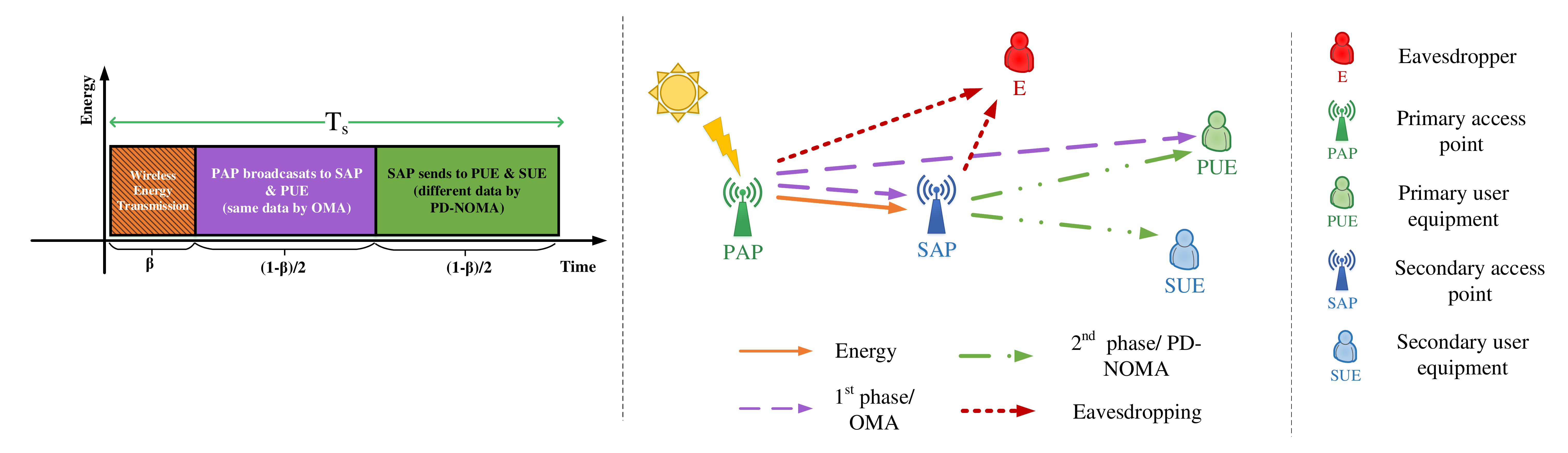} %\vspace{-1cm}
			\caption{Network model for the EH secure CR network with PD-NOMA technique, where the PAP harvests energy from renewable source and the SAP absorbs energy based on time switching technique and use PD-NOMA protocol. }\label{Fig:SystemModelNOMA}%\vspace{-.5cm}
		\end{center}%\vspace{-0.7cm}
	\end{figure}

	We consider an {EH} downlink cooperative secure PD-NOMA network with primary transmission which works as an overlay network, secondary transmission that works as an underlay network, and an eavesdropper\footnote{Here, we suppose that the eavesdropper is an idle user of the network which does not have permission to listen to some messages.}. The architecture of the system model is illustrated in Fig.~\ref{Fig:SystemModelNOMA}.
	
	The primary network consists of a PAP as a transmitter equipped with a finite capacity battery with the capability of harvesting energy from green energies such as solar energy and wind power and storing surplus of energy in the battery and a PUE which is served by the PAP. The secondary network underlays the primary one and uses the bandwidth of the primary network with agreement on helping to forward the PAP data toward its receiver. The secondary network consists of one SAP as a transmitter and a secondary user equipment (SUE) as a receiver. The SAP harvests energy from the RF signals transmitted by the PAP in the TS mode and uses this energy to decode-and-forward the signal of the PAP besides sending its own data. The SAP is equipped with a finite battery to save surplus energy units for future usage. The malicious user in this network tries to eavesdrop data of all transmitters through the direct link and the cooperative link. Moreover, each node is equipped with a single antenna.  
	
	Let $T_D$ be the number of time slots over which the communication system is studied  and the length of each time slot is $T_s$. Moreover, we assume that the CSI remains constant during each time slot $t_d$ and changes independently from one time slot to other time slot. 	As shown in Fig.\ref{Fig:SystemModelNOMA},  OMA and PD-NOMA techniques are adopting at the PAP and the SAP, respectively, for data transmission. %	In addition, SAP utilizes the TS protocol to absorb energy from the PAP. We suppose that each time slot is divided into three phases.
	{In} the first phase, the PAP sends energy with power $p_{pp}(t_d)$ with duration $\beta\in[0,1]$ {to the SAP}. $\beta=1$ means that there is no time for data transmission and $\beta=0$ means that the system considers that there is no need for energy transmission. In this phase, the SAP harvests {energy from the RF signal and employs it t}o forward the PAP data to the PUE, as well as its own information to the SUE. The harvested energy at the SAP can be expressed as follows:
	\begin{align}
		\label{Energy_harvesing_WET}E_s^{}(t_d)&=T_s\beta\eta_{2,s}  p_{pp}(t_d) h_{\text{PAP}\rightarrow \text{SAP}}(t_d),
	\end{align}
	where $h_{\text{PAP}\rightarrow \text{SAP}}(t_d)$ is  the {channel gain} from the PAP to the PUE at time slot $t_d$ and $\eta_{2,s}\in (0,1] $ is the energy conversion efficiency at the SAP. 	During the second phase, the PAP broadcasts its message based on  OMA technique.	The signal to noise ratio (SNR) at the PUE is calculated as follows:
	\begin{align}
		\label{SNR_PUE_from_PAP} \gamma_{\text{PAP}\rightarrow\text{PUE}}^\text{}(t_d)=\frac{p_{pp}(t_d)h_{\text{PAP}\rightarrow\text{PUE}}(t_d)}{\sigma^2_{pp}},
	\end{align}
	where  $\sigma^2_{pp}$ is the variance of additive Gaussian white noise (AWGN) at the PUE. The received data rate in {(bit/s/Hz)} at the SAP is given by
	\begin{align}
%		\label{TS_OMA_SNR_SP}\gamma_{\text{PAP}\rightarrow \text{SAP}}^{}(t_d)&=\frac{p_{pp}(t_d)h_{\text{PAP}\rightarrow \text{SAP}}(t_d)}{\sigma_{ps}^2},\\
		r_{\text{PAP}\rightarrow \text{SAP}}^{}(t_d)&=\frac{(1-\beta)}{2}\log_2\Big(1+\frac{p_{pp}(t_d)h_{\text{PAP}\rightarrow \text{SAP}}(t_d)}{\sigma_{ps}^2}\Big),
	\end{align}
	where $h_{\text{PAP}\rightarrow \text{SAP}}(t_d)$ is the channel gain from the PAP to the SAP and  $\sigma^2_{ps}$ is the variance of AWGN at the SAP. The SNR at the eavesdropper can be formulated as 
	\begin{align}\label{SNR_PAP_E}
		\gamma_{\text{PAP}\rightarrow E}(t_d)=\frac{p_{pp}(t_d)h_{\text{PAP}\rightarrow E}(t_d)}{\sigma_{pe}^2},
		%	r_{\text{PAP}\rightarrow E}(t_d)=\frac{T_s}{3}\log_2\big(1+\gamma_{\text{PAP}\rightarrow E}(t_d)\big)
	\end{align} 
	where $h_{\text{PAP}\rightarrow E}(t_d)$ is channel gain from the PAP to the eavesdropper and   $\sigma_{pe}^2$ is the variance of AWGN at the eavesdropper side. 
	{In} the third phase, the SAP  decodes the received signal from the PAP and by utilizing  PD-NOMA technique, it simultaneously sends the decoded signal and its own message towards the PUE and the SUE, respectively.
	
	Based on the relation between the {channel gain} from the SAP to the PUE, i.e., $h_{\text{SAP}\rightarrow \text{PUE}}(t_d)$, and the channel gain from the SAP to the SUE, i.e., $h_{\text{SAP}\rightarrow \text{SUE}}(t_d)$, there are two scenarios. In the first scenario, if $h_{\text{SAP}\rightarrow \text{PUE}}(t_d)\geq h_{\text{SAP}\rightarrow \text{SUE}}(t_d)$,
	%the transmission power should be $p_{ss}(t_d)\geq p_{sp}(t_d)$ 
	 the successive interference cancellation (SIC) is implemented at the PUE, and the SUE suffers from interference. Hence, the SNR at the PUE after removing the interference can be written as follows:
	\begin{align}\label{PS_NOMA_SNR_NOMA_SEN_I_sp}
		\gamma_{\text{SAP}\rightarrow \text{PUE}}^{\text{SCN I}} (t_d)=\frac{p_{sp}(t_d)h_{\text{SAP}\rightarrow \text{PUE}}(t_d)}{\sigma_{sp}^2},
	\end{align}
	where $\sigma^2_{sp}$ is {the} variance of AWGN, then the SNR at the SUE can be shown as follows:
	\begin{align}\label{PS_NOMA_SNR_NOMA_SEN_I_ss}
		\gamma_{\text{SAP}\rightarrow \text{SUE}}^{\text{SCN I}} (t_d)=\frac{p_{ss}(t_d)h_{\text{SAP}\rightarrow \text{SUE}}(t_d)}{p_{sp}(t_d)h_{\text{SAP}\rightarrow\text{SUE}}(t_d)+\sigma_{ss}^2},
	\end{align}
	where $\sigma^2_{ss}$ is {the} variance of AWGN.	{In the second scenario}, if $h_{\text{SAP}\rightarrow \text{SUE}} (t_d)> h_{\text{SAP}\rightarrow \text{PUE}}(t_d)$, the SIC is implemented by the SUE and the PUE suffers from  PD-NOMA interference.
	% In addition, the amount of transmit power should satisfy  $p_{ss}(t_d) < p_{sp}(t_d)$.
	 The throughput at the SUE is expressed as 
	\begin{align}
		\label{PS_NOMA_SNR_SEN_II_ss}r_{\text{SAP}\rightarrow \text{SUE}}^{\text{SCN II}}(t_d)&=\frac{(1-\beta)}{2}\log_2\Big(1+\frac{p_{ss}(t_d)h_{\text{SAP}\rightarrow \text{SUE}}(t_d)}{\sigma_{ss}^2}\Big),
	\end{align}
	and the data rate at the PUE side can be formulated as follows:
	\begin{align}\label{PS_NOMA_SNR_SEN_II_sp}
		r_{\text{SAP}\rightarrow \text{PUE}}^{\text{SCN II}}(t_d)=\frac{(1-\beta)}{2}\log_2\Big(1+\frac{p_{sp}(t_d)h_{\text{SAP}\rightarrow \text{PUE}}(t_d)}{p_{ss}(t_d)h_{{\text{SAP}\rightarrow \text{PUE}}}(t_d)+\sigma_{sp}^2}\Big).
	\end{align}

%	Based on these two scenarios, the throughput of SUE can be formulated as follows:
%	\begin{align}
%		r_{\text{SAP}\rightarrow \text{SUE}}^{\text{S}}(t_d)=\frac{(1-\beta)}{2}\log_2\big(1+\gamma_{\text{SAP}\rightarrow \text{SUE}}^{\text{NOMA},\text{S}}(t_d)\big),
%	\end{align}
%	where $\text{S}\in [\text{SCN I}, \text{SCN II}]$.

	Here, we consider {the {worst} case for eavesdropper  where it uses  strong multiuser detection techniques and can successfully cancel the interference signal  from the PAP and the SAP \cite{Lv1,Liu2}}. The received SNR at the eavesdropper at this phase is
	\begin{align}\label{SNR_SAP1_E}
		\gamma_{\text{SAP}1\rightarrow {E} } (t_d)&=\frac{h_{\text{SAP}\rightarrow E}(t_d) p_{sp}(t_d)} {\sigma_{se}^2}.
	\end{align}
	
	At the end of second phase, the PUE catches information from the PAP and the SAP, and decodes the received massages jointly by using the maximum ratio combining (MRC). Therefore, the data rate in ({bit/s/Hz}) based on the MRC in the PUE can be described as
	\begin{align}
%		\label{PS_NOMA_SNR_MRC_p}\gamma_{\text{MRC-P}}^{\text{S}}(t_d) &=\gamma_{\text{PAP}\rightarrow \text{PUE}}(t_d)+\gamma_{\text{SAP}\rightarrow \text{PUE}}^{\text{S}}(t_d),\\
		r_{\text{MRC-P}}^{\text{S}}(t_d)&=\frac{(1-\beta)}{2}\log_2\big(1+\gamma_{\text{PAP}\rightarrow \text{PUE}}(t_d)+\gamma_{\text{SAP}\rightarrow \text{PUE}}^{\text{S}}(t_d)\big),
	\end{align}
	where $\text{S}\in [\text{SCN I}, \text{SCN II}]$.
	
	The throughput related to the PUE's data at  the eavesdropper can be formulated as follows:
	\begin{align}\label{MRC_E}
%		\gamma^\text{MRC}_{P\rightarrow E}  (t_d)&= \gamma_{\text{PAP}\rightarrow E}(t_d)+\gamma_{\text{SAP}1\rightarrow {E} } (t_d),\\
		r_{\text{MRC-E} } (t_d)&=\frac{1-\beta}{2}\log_2\big(1+\gamma_{\text{PAP}\rightarrow E}(t_d)+\gamma_{\text{SAP}1\rightarrow {E} } (t_d)\big).
	\end{align}
	The data rate in ({bit/s/Hz}) related to the SUE's data at the eavesdropper can be formulated as follows:
	{\begin{align}
%		\label{SNR_SUE_E}\gamma_{\text{SAP}2\rightarrow E}(t_d)&=\frac{p_{ss}(t_d)h_{\text{SAP}\rightarrow E}(t_d)}{\sigma_{ss}^2},\\
		r^{}_{\text{SAP2}\rightarrow E }(t_d)&=\frac{1-\beta}{2}\log_2\Big(1+\frac{p_{ss}(t_d)h_{\text{SAP}\rightarrow E}(t_d)}{\sigma_{ss}^2}\Big).
	\end{align}}
		We {suppose} that the harvested energy will {be stored} in the battery and each transmitter only uses the stored energy in their batteries. Hence, at the beginning of time slot $t_d$, the remaining energy in the batteries can be formulated as 
	\begin{align}
		B_p^{}(t_d) &= \min\{B_{\max,p}, B_p^{}(t_d-1) + \eta_{1,p}E_{p}^\text{RE}(t_d-1) - T_s\big(\beta+\frac{(1-\beta)}{2}) p_{pp}(t_d-1)-E_{\text{P}}^{\text{Cons}}  \}, \\
		B_s^{}(t_d)&= \min\{ B_{\max,s}, B_s^{}(t_d-1) + {\eta_{1,s}E_{s}(t_d)} - T_s\frac{(1-\beta)}{2} \big(p_{sp}(t_d-1)+p_{ss}(t_d-1)\big)-E_{\text{S}}^{\text{Cons}}  \},
	\end{align} 
	where $B_p^{}(t_d-1)$ and $B_s^{}(t_d-1)$ are the energy value of batteries of the PAP and the SAP at the beginning of previous time slot, and $B_p^{}(0)=B_{p,0}$ and $B_s^{}(0)=B_{s,0}$. Let {{${E_p^{\text{RE}}(t_d)}$}} denote the amount of energy received at the PAP at time slot $t_d$.   $E_{\text{P}}^{\text{Cons}}$ and {$E_{\text{S}}^{\text{Cons}}$} are the constant energy consumption {for the operation of the PAP and the SAP.} Moreover, $B_{\max,p}$ and $B_{\max,s}$ are the maximum capacity of batteries for the PAP and the SAP, respectively.  $\eta_{1,p}\in  (0, 1]$ and $\eta_{1,s} \in (0, 1]$ are the storage efficiencies of the PAP and the SAP. It is worth mentioning that we assume the SAP can use the harvested energy from the PAP in the same time slot since the SAP will begin to transmit if and only if the signal reception from the PAP has ended. By the way, the PAP uses the harvested energy only at the next time slot when energy units have  received completely.

	The total secrecy rate of the network at the end of each time slot can be formulated as follows:
	\begin{align}
		R^{}(t_d) &= \Big[\min\big(r^{}_{\text{PAP}\rightarrow \text{SAP}}(t_d),r_{\text{MRC-P}}^{\text{S}}(t_d)\big)-r_{\text{MRC-E} } (t_d)\Big]^+ + \Big[r_{\text{SAP}\rightarrow \text{SUE}} ^{\text{S}}(t_d)-r^{}_{\text{SAP2}\rightarrow E }(t_d)\Big]^+,
	\end{align}
	%where 
	%\begin{align}
	%	r_p(t_d)&=\Big[\min\big(r^{}_{\text{PAP}\rightarrow \text{SAP}}(t_d),r_{\text{MRC-P}}^{\text{S}}(t_d)\big)-r_{\text{MRC-E} } (t_d)\Big]^+,\\
	%	r_{s}(t_d)&=\Big[r_{\text{SAP}\rightarrow \text{SUE}} ^{\text{S}}(t_d)-r^{}_{\text{SAP2}\rightarrow E }(t_d)\Big]^+.
	%\end{align}
	and the total energy consumption at each time slot can be {computed} as follows:
	\begin{align}
		&E^{}(t_d) =T_s \nu  \Big( \big(\beta+\frac{(1-\beta)}{2}\big) p_{pp}(t_d)+ \frac{(1-\beta)}{2}\big(p_{sp}(t_d)+p_{ss}(t_d)\big) \Big)-\nu E^{}_s(t_d)+ E_c.
	\end{align}
	
	\subsection {Uncertainty and Robustness}
	The uncertainty can be classified into two categories as aleatoric and epistemic which arise from stochastic behavior and parameter estimation, respectively \cite{salahdine2017techniques}. The aleatoric is a statistical and random uncertainty and cannot be predicted by collecting more information and knowledge. On the other hand, the epistemic uncertainty can be modeled by using probability distribution functions and diminished through collecting more information about the uncertain system \cite{salahdine2017techniques}. In the CR networks, we are dealing with the epistemic uncertainty, therefore it can be handled by using methods which collects previous information of the uncertain system \cite{salahdine2017techniques}. In the following, we investigate the uncertainties about channel gains and battery energy values. 
	\subsubsection{Channel Uncertainty}
	Here, we suppose that the perfect  CSI between the transmitters and the receivers is not available. This uncertainty comes from the disability of the network in catching the appropriate value of channel gain, due to the error in the feedback channels and/or out-date channel value \cite{setoodeh2009robust, salahdine2017techniques}. The uncertainty of channel gains can be represented by the summation of estimated channel value and additive error as follows:
	\begin{align}
		h_{\text{PAP}\rightarrow\text{PUE}}(t_d)&=\bar {h}_{\text{PAP}\rightarrow\text{PUE}}(t_d)+\epsilon_{h_{\text{PAP}\rightarrow\text{PUE}}(t_d)},\\
		h_{\text{PAP}\rightarrow\text{SAP}}(t_d)&=\bar {h}_{\text{PAP}\rightarrow\text{SAP}}(t_d)+\epsilon_{h_{\text{PAP}\rightarrow\text{SAP}}(t_d)},\\
		h_{\text{SAP}\rightarrow \text{PUE}}(t_d)&=\bar {h}_{\text{SAP}\rightarrow \text{PUE}}(t_d)+\epsilon_{h_{\text{SAP}\rightarrow \text{PUE}}(t_d)},\\
		h_{\text{SAP}\rightarrow \text{SUE}}(t_d)&=\bar {h}_{\text{SAP}\rightarrow \text{SUE}}(t_d)+\epsilon_{h_{\text{SAP}\rightarrow \text{SUE}}(t_d)},
	\end{align}
	where $\bar{h}_{[\cdot]}$ and $\epsilon_{[\cdot]}$ are the estimated and error values of channel gain, respectively, and  $h_{[\cdot]}(t_d)$ {is} the exact values of the channels. Moreover, we define {the} uncertainty region for each channel gain as the distance from the actual and the estimated values of that channel as follows:
	{\begin{align}
		&\mathcal{R}_{h_{\text{PAP}\rightarrow\text{PUE}}}=\{h_{\text{PAP}\rightarrow\text{PUE}}(t_d) |~| h_{\text{PAP}\rightarrow\text{PUE}}(t_d)-\bar{h}_{\text{PAP}\rightarrow\text{PUE}}(t_d) | \leq \delta _{h_{\text{PAP}\rightarrow\text{PUE}}}\},\\
		&\mathcal{R}_{h_{\text{PAP}\rightarrow\text{SAP}}}=\{h_{\text{PAP}\rightarrow\text{SAP}}(t_d) |~| h_{\text{PAP}\rightarrow\text{SAP}}(t_d)-\bar{h}_{\text{PAP}\rightarrow\text{SAP}}(t_d) | \leq \delta _{h_{\text{PAP}\rightarrow\text{SAP}}}\},\\
		&\mathcal{R}_{h_{\text{SAP}\rightarrow \text{PUE}}}=\{h_{\text{SAP}\rightarrow \text{PUE}}(t_d) |~| h_{\text{SAP}\rightarrow \text{PUE}}(t_d)-\bar{h}_{\text{SAP}\rightarrow \text{PUE}}(t_d) | \leq \delta _{h_{\text{SAP}\rightarrow \text{PUE}}}\},\\
		&\mathcal{R}_{h_{\text{SAP}\rightarrow \text{SUE}}}=\{h_{\text{SAP}\rightarrow \text{SUE}}(t_d) |~| h_{\text{SAP}\rightarrow \text{SUE}}(t_d)-\bar{h}_{\text{SAP}\rightarrow \text{SUE}}(t_d) | \leq \delta _{h_{\text{SAP}\rightarrow\text{SUE}}}\},
	\end{align}}
	where {$\delta_{[\cdot]}$} is bound on the uncertainty region and $|\cdot|$ is the absolute value. {The channel gains from the PAP to the eavesdropper and the SAP to the eavesdropper are not observable for the PAP and the SAP.} 
	
	\subsubsection{Battery Level Uncertainty}
	We consider that the {amount of remaining energy measured from battery is not accurate}. {The uncertainty in  energy level of battery is considered by \cite{Michelusi1} and \cite{Yang1}}. The uncertainty comes from the battery dynamics and quantization of sensor value. Usually the embedded batteries in these kinds of EH devices have a very low capacity and estimation of accurate values of batteries is costly or impractical \cite{Michelusi1}. We denote the uncertainty of battery value in the PAP and the SAP by the estimated battery value plus the additional error which can be formulated as follows:
	{\begin{align}
		B_p(t_d)&=\bar{B}_p(t_d)+\epsilon_{B_p(t_d)},\\
		B_s(t_d)&=\bar{B}_s(t_d)+\epsilon_{B_s(t_d)},
	\end{align}}        
	where $\bar{B}_{[\cdot]}(t_d)$ and $\epsilon_{[\cdot]}$ are the estimated and error values of batteries, respectively. Moreover, the uncertainty region for the batteries value can be formulated as follows:
	\begin{align}
		\mathcal{R}_{B_p}&=\{B_p(t_d) |~| B_p(t_d)-\bar{B}_{p}(t_d) | \leq \delta _{B_p}\},\\	
		\mathcal{R}_{B_s}&=\{B_s(t_d) |~| B_s(t_d)-\bar{B}_{s}(t_d) | \leq \delta _{B_s}\}.
	\end{align}

	\subsection{Optimization problem of robust resource allocation}	
The robust resource allocation problem with objective of maximizing the PFEE is considered to find near optimal transmit power and the TS ratio in the TS-NOMA system model, and can be formulated as follows:

{
	\begin{subequations}\label{O_TS_WET_NOMA}
		\begin{align}
			&\mathcal{O}^{\text{TS-NOMA}}:\nonumber\\
			\label{O_TS_WET_NOMA_OF}\max_{\boldsymbol{p}_{pp},\boldsymbol{p}_{sp},\boldsymbol{p}_{ss},\beta}~~&\sum\limits_{t_d=1}^{T_D} \min_{\underline{\boldsymbol{h}}\in \underline{\boldsymbol{\mathcal{R}}}} \log_2\Big(\frac{R^{}(t_d)}{E^{}(t_d)}\Big),\\ 
			\text{s.t.}~~	\label{O_TS_WET_NOMA_C1} \text{C1: }&\min_{B_p^{}(t_d) \in \mathcal{R}_{B_p^{}}}B_p^{}(t_d) -T_s\Big(\beta+\frac{(1-\beta)}{3}\Big) p_{pp}(t_d) \geq 0, ~~~~~~~~~~~~~~~~~\forall t_d\in\{1,\ldots,T_D\},\\
			\label{O_TS_WET_NOMA_C2}\text{C2: }&\min_{B_s^{}(t_d) \in \mathcal{R}_{B_s^{}}} B_s^{}(t_d)+	\min_{h_{\text{PAP}\rightarrow \text{SAP}(t_d)}\in\mathcal{R}_{h_{\text{PAP}\rightarrow \text{SAP}}}}\eta_{1,s}E^{}_s(t_d)\nonumber\\
			&-T_s\frac{ (1-\beta)}{2}\big(p_{ss}(t_d)+p_{sp}(t_d)\big)\geq 0 ,~~~~~~~~~~~~~~~~~~~~~~~~~~~~~~~~~\forall t_d\in\{1,\ldots,T_D\},\\	
	\label{O_TS_WET_NOMA_C3}	\text{C3: }&\max_{B_p^{}(t_d) \in \mathcal{R}_{B_p^{}}} B_p^{}(t_d) + \eta_{1,p}E^{\text{RE}}_p(t_d) -T_s\Big(\beta+\frac{(1-\beta)}{2}\Big) p_{pp}(t_d) \leq B_{\max,p} ,~~\forall t_d\in\{1,\ldots,T_D\},\\
			\label{O_TS_WET_NOMA_C4}\text{C4: }& \max_{B_s^{}(t_d) \in \mathcal{R}_{B_s^{}}}B_s^{}(t_d)+\max_{h_{\text{PAP}\rightarrow \text{SAP}(t_d)}\in\mathcal{R}_{h_{\text{PAP}\rightarrow \text{SAP}}}}\eta_{1,s}\sum\limits_{i=t_d}^{t_d+1} E^{}_s(i) \nonumber\\
			&-\frac{T_s(1-\beta)}{2}\big(p_{ss}(t_d)+p_{sp}(t_d)\big) \leq B_{\max,s}  ,~~~~~~~~~~~~~~\forall t_d\in\{1,\ldots,T_D\},\\
			\label{O_TS_WET_NOMA_C5}\text{C5: }&  \sum\limits_{t_d=1}^{T_D}\min\Big( \min_{
					h_{\text{PAP}\rightarrow \text{SAP}}(t_d)\in \mathcal{R}_{h_{\text{PAP}\rightarrow \text{SAP}}}
				} r_{\text{PAP}\rightarrow\text{SAP}}^{}(t_d), \min_{\underline{\underline{\boldsymbol{h}}}\in \underline{\underline{\boldsymbol{\mathcal{R}}}}}r^{\text{S}}_{\text{MRC-P}} (t_d) \Big)\geq R^{\min},\\
			\label{O_TS_WET_NOMA_C6}\text{C6: }&p_{pp}(t_d)\leq P_{p}^{\max},~~~~~~~~~~~~~~~~~~~~~~~~~~~~~~~~~~~~~~~~~~~~~~~~~~~~ \forall t_d\in\{1,\ldots,T_D\},\\
			\label{O_TS_WET_NOMA_C7}\text{C7: }&p_{ss}(t_d) + p_{sp}(t_d) \leq P_{s}^{\max},~~~~~~~~~~~~~~~~~~~~~~~~~~~~~~~~~~~~~~~~~\forall t_d\in\{1,\ldots,T_D\},\\
			\label{O_TS_WET_NOMA_C8}\text{C8: }&0<\beta<1,\\
			\label{O_TS_WET_NOMA_C9}\text{C9: }& h_{\text{PAP}\rightarrow\text{PUE}}(t_d)\in \mathcal{R}_{h_{\text{PAP}\rightarrow\text{PUE}}},  h_{\text{PAP}\rightarrow\text{SAP}}(t_d)\in \mathcal{R}_{h_{\text{PAP}\rightarrow\text{SAP}}},\nonumber \\
			&h_{\text{SAP}\rightarrow\text{PUE}}(t_d)\in \mathcal{R}_{h_{\text{SAP}\rightarrow\text{PUE}}}, h_{\text{SAP}\rightarrow\text{SUE}}(t_d)\in \mathcal{R}_{h_{\text{SAP}\rightarrow\text{SUE}}}, \\
			\label{O_TS_WET_NOMA_C10}\text{C10: }&B_{p}^{}(t_d)\in \mathcal{R}_{B_{p}^{}}, B_{s}^{}(t_d)\in \mathcal{R}_{B_{s}^{}},
		\end{align}
	\end{subequations}}
	where %$B_{p,0}$ and $B_{s,0}$ are the initial energy at the PAP and the SAP, respectively, 
	$\boldsymbol{p}_{pp}=\{p_{pp}(t_d), t_d\in T_D  \}$, $\boldsymbol{p}_{sp}= \{p_{sp}(t_d), t_d\in T_D  \}$, $\boldsymbol{p}_{ss}=\{p_{ss}(t_d), t_d\in T_D  \}$, 
	$B_{\max,p}$ and $B_{\max,s}$ are the maximum capacity of battery at the PAP and the SAP, respectively. Let $\underline{\boldsymbol{h}} =[h_{\text{PAP}\rightarrow \text{PUE}}(t_d)$ $,h_{\text{PAP}\rightarrow \text{SAP}}(t_d),h_{\text{SAP}\rightarrow \text{PUE}}(t_d),h_{\text{SAP}\rightarrow \text{SUE}}(t_d)]^T$, $\underline{\boldsymbol{\mathcal{R}}} = \mathcal{R}_{h_{\text{PAP}\rightarrow \text{PUE}}}\times\mathcal{R}_{h_{\text{PAP}\rightarrow \text{SAP}}}\times\mathcal{R}_{h_{\text{SAP}\rightarrow \text{PUE}}}\times\mathcal{R}_{h_{\text{SAP}\rightarrow \text{SUE}}}$, $\underline{\underline{\boldsymbol{h}}} =[$ $h_{\text{PAP}\rightarrow \text{PUE}}(t_d)$ $,h_{\text{SAP}\rightarrow \text{PUE}}(t_d),h_{\text{SAP}\rightarrow \text{SUE}}(t_d)]^T$, and $\underline{\underline{\boldsymbol{\mathcal{R}}}} = \mathcal{R}_{h_{\text{PAP}\rightarrow \text{PUE}}}\times\mathcal{R}_{h_{\text{SAP}\rightarrow \text{PUE}}}\times\mathcal{R}_{h_{\text{SAP}\rightarrow \text{SUE}}}$. The minimum {required} rate of the PUE is determined by $R_{\min,p}$. Moreover, $P_{p}^{\max}$ and $P_{s}^{\max}$ are the maximum transmit power of the PAP and the SAP, respectively. \eqref{O_TS_WET_NOMA_C1} and \eqref{O_TS_WET_NOMA_C2} are causality constraints at the PAP and the SAP, respectively. \eqref{O_TS_WET_NOMA_C3} and \eqref{O_TS_WET_NOMA_C4} are battery over flow constraints at the PAP and the SAP, respectively. The constraint \eqref{O_TS_WET_NOMA_C5} guarantees that the minimum required data rate of the PUE is satisfied. \eqref{O_TS_WET_NOMA_C6} and \eqref{O_TS_WET_NOMA_C7} ensure that the total transmit power at each phase should not exceed the maximum power budget. \eqref{O_TS_WET_NOMA_C8} shows the range of TS ratio. Finally, \eqref{O_TS_WET_NOMA_C9} and \eqref{O_TS_WET_NOMA_C10}  show the uncertainty regions related to channel gains and batteries, respectively. 
	
		\subsection{Worst case scenario}
	In the practice, the underestimation of channel conditions or other parameters such as energy value stored in the battery is inevitable. Here, we expand our problem into the worst case robust optimization problem. The worst case counterpart of our optimization problem can be reformulated as

	\begin{subequations}\label{O_TS_WET_NOMA2}
		\begin{align}\label{opt_2}
			&\max_{\boldsymbol{p}_{pp},\boldsymbol{p}_{sp},\boldsymbol{p}_{ss},\beta,\boldsymbol{u},\boldsymbol{v}}~~\sum\limits_{t_d=1}^{T_D}\log_2\Big(\frac{\bar{\bar{R}}_\text{W}(t_d)}{\bar{\bar{E}}_\text{W}(t_d)}\Big),\\
			\text{s.t.~~}&	\text{C6, C7, C8},\\
			&B_p^{}(t_d)-\delta_{B_p^{}}  -T_s\Big(\beta+\frac{(1-\beta)}{2}\Big)p_{pp}(t_d)\geq 0,~~~~~~~~~~~~~~~~~~~~~~~~\forall t_d\in\{1,\ldots,T_D\},\\
			&B_s^{}(t_d)-\delta_{B_s^{}}+{T_s}\beta	\eta_{1,s}\eta_{2,s}p_{pp}(t_d) \big(\bar{h}_{\text{PAP}\rightarrow \text{SAP}}(t_d)-\delta_{h_{\text{PAP}\rightarrow\text{SAP}}}\big)-\nonumber\\
			&T_s\frac{(1-\beta)}{2}\big(p_{ss}(t_d)+p_{sp}(t_d)\big)\geq 0 ,~~~~~~~~~~~~~~~~~~~~~~~~~~~~~~~~~~~~~~\forall t_d\in\{1,\ldots,T_D\},\\
			& B_p^{}(t_d)+\delta_{B_p^{}} + \eta_{1,p}E^{\text{RE}}_p(t_d) -T_s\Big(\beta+\frac{(1-\beta)}{2}\Big)p_{pp}(t_d) \leq B_{\max,p} ,~~\forall t_d\in\{1,\ldots,T_D\},\\
			&B_s^{}(t_d)+\delta_{B_s^{}} +{T_s}\beta\eta_{1,s}\eta_{2,s}\sum\limits_{i=t_d}^{t_d+1}p_{pp}(i) \big(\bar{h}_{\text{PAP}\rightarrow \text{SAP}}(i)-\delta_{h_{\text{PAP}\rightarrow\text{SAP}}}\big) \nonumber\\
			&-T_s\frac{(1-\beta)}{2} \big(p_{ss}(t_d)+p_{sp}(t_d)\big) \leq B_{\max,s}  ,~~~~~~~~~~~~~~~~~~~~~~~~~~~~~\forall t_d\in\{1,\ldots,T_D\},\\
			& \sum\limits_{t_d=1}^{T_D} \bar{\bar{R}}^{\min}_\text{W}(t_d) \geq R_{\min,p},~~~~~~~~~~~~~~~~~~~~~~~~~~~~~~~~~~~~~~~~~~~~~~~~~~~\forall t_d\in\{1,\ldots,T_D\},
		\end{align}
	\end{subequations}
	where
	\begin{align}
		&\bar{\bar{R}}_\text{W}(t_d)=
		\frac{(1-\beta)}{2}\Big[\bar{\bar{r}}_{\text{SAP}\rightarrow \text{SUE}} ^{\text{S}}(t_d)-\log_2\Big(1+\frac{p_{ss}(t_d)\big(\bar{h}_{\text{SAP}\rightarrow E}(t_d)+\delta_{h_{\text{SAP}\rightarrow E}}\big) } {\sigma_{se}^2}\Big)\Big]^+ +
		\\
		&\frac{(1-\beta)}{2}\Big[\bar{\bar{R}}^{\min}_{\text{W}}(t_d)-\log_2\Big(1+\frac{p_{pp}(t_d)\big(\bar{h}_{\text{PAP}\rightarrow E}(t_d)+\delta_{h_{\text{SAP}\rightarrow E}}\big)}{\sigma_{pe}^2}+\frac{p_{sp}(t_d)\big(\bar{h}_{\text{SAP}\rightarrow E}(t_d)+\delta_{h_{\text{SAP}\rightarrow E}}\big) } {\sigma_{se}^2} \Big)\Big]^+,\nonumber\\
		&\bar{\bar{E}}^{}_\text{W}(t_d)={T_s}\nu \Big( \big(\frac{(1-\beta)}{2}+\beta\big)p_{pp}(t_d)+\frac{(1-\beta)}{2} p_{sp}(t_d) +\frac{(1-\beta)}{2}p_{ss} (t_d) \Big) -\\
		&~~~~~~~~~~~~~{T_s}\nu  \beta  \eta_{2,s}  p_{pp}(t_d) \big(\bar{h}_{\text{PAP}\rightarrow\text{SAP}}(t_d)-\delta_{h_{\text{PAP}\rightarrow\text{SAP}}}\big)  + E_c,\nonumber
			\end{align}
	\begin{align}
		&\bar{\bar{R}}^{\min}_\text{W}(t_d)=\min\big\{\log_2\Big(1+\frac{p_{pp}(t_d)\big(\bar{h}_{\text{PAP}\rightarrow\text{SAP}}(t_d)-\delta_{h_{\text{PAP}\rightarrow\text{SAP}}}\big)}{\sigma_{ps}^2}\Big),\bar{\bar{r}}_{\text{MRC-P}}^{\text{S}} (t_d)\big\},
	\end{align}
	and
{	\begin{align}
		\bar{\bar{r}}_{\text{SAP}\rightarrow \text{SUE}} ^{\text{SCN I}}(t_d)&=\log_2\Big(1+\frac{p_{ss}(t_d)\big(\bar{h}_{\text{SAP}\rightarrow \text{SUE}}(t_d)-\delta_{h_{\text{SAP}\rightarrow \text{SUE}}}\big)}{p_{sp}(t_d)\big(\bar{h}_{\text{SAP}\rightarrow\text{SUE}}(t_d)-\delta_{h_{\text{SAP}\rightarrow \text{SUE}}}\big)+\sigma_{ss}^2}\Big),\\
		\bar{\bar{r}}_{\text{SAP}\rightarrow \text{SUE}} ^{\text{SCN II}}(t_d) &= \log_2\Big(1+\frac{p_{ss}(t_d)\big(\bar{h}_{\text{SAP}\rightarrow \text{SUE}}(t_d)-\delta_{h_{\text{SAP}\rightarrow \text{SUE}}}\big)}{\sigma_{ss}^2} \Big),\\
		\bar{\bar{r}}_{\text{MRC-P}}^{\text{SCN I}}(t_d)&=\log_2\Big( 1+ \frac{p_{pp}(t_d)\big(\bar{h}_{\text{PAP}\rightarrow\text{PUE}}(t_d)-\delta_{h_{\text{PAP}\rightarrow\text{PUE}}}\big)}{\sigma^2_{pp}} + \frac{p_{sp}(t_d)\big(\bar{h}_{\text{SAP}\rightarrow \text{PUE}}(t_d)-\delta_{h_{\text{SAP}\rightarrow \text{PUE}}}\big)}{\sigma_{sp}^2} \Big),\\
		\bar{\bar{r}}_{\text{MRC-P}}^{\text{SCN II}}(t_d)&=\log_2\Big( 1+ \frac{p_{pp}(t_d)\big(\bar{h}_{\text{PAP}\rightarrow\text{PUE}}(t_d)-\delta_{h_{\text{PAP}\rightarrow\text{PUE}}}\big)}{\sigma^2_{pp}} + \frac{p_{sp}(t_d)\big(\bar{h}_{\text{SAP}\rightarrow \text{PUE}}(t_d)-\delta_{h_{\text{SAP}\rightarrow \text{PUE}}}\big)}{p_{ss}(t_d)\big(\bar{h}_{{\text{SAP}\rightarrow \text{PUE}}}(t_d)-\delta_{h_{\text{SAP}\rightarrow \text{PUE}}}\big)+\sigma_{sp}^2} \Big).
	\end{align}}
	by taking derivation from (34) and (37) respect to the $h_{\text{SAP}\rightarrow \text{SUE}}$ and $h_{\text{SAP}\rightarrow\text{PUE}}$, respectively, it is provable that  both of them are increasing function respect to their own derivation parameters. Hence, the minimum value of these equations can be achievable by minimum value of $h_{\text{SAP}\rightarrow \text{SUE}}$ and $h_{\text{SAP}\rightarrow \text{PUE}}$. 
	
	\subsection{Stochastic  scenario}

	Uncertainty in the amount of channel gain and energy level in battery has stochastic behavior. By considering this nature, the constraints with uncertain parameters can be rewritten as a violation probability with a predefined threshold. In our resource allocation problems, firstly we rewrite the resource allocation problem as the violation probability, then we use Bernstein approximation scheme  and calculate the close form for each constraint.

	The uncertainty is due to the difference between real value and estimated value, the error value, which has the stochastic nature.  
	By considering the stochastic nature of the error value in the uncertain parameters, the resource allocation problem $\mathcal{O}^{\text{TS-NOMA}}$ in the form of chance constrained can be formulated as follows:
	\begin{subequations}\label{Ps_OMA_stochastic}
		\begin{align}
			\max_{\boldsymbol{p}_{pp},\boldsymbol{p}_{sp},\boldsymbol{p}_{ss},\beta,\boldsymbol{u},\boldsymbol{v}}~~&\sum\limits_{t_d=1}^{T_D}\log_2\Big(\frac{u(t_d)}{v(t_d)}\Big),\\
			\text{s.t.~~}& \text{C6, C7, C8, C9, C10},\nonumber\\
			\label{probab_c1}&\text{Pr}\Big({R}_{}(t_d) \leq u(t_d) \Big) \leq \xi_{\text{Obj}_\text{Nom}},~~~~~~~~~~~~~~~~~~~~~~~~~~~~~~~~~~~\forall t_d\in\{1,\ldots,T_D\},\\
			\label{probab_c2}&\text{Pr}\Big({E}_{}(t_d) \geq v(t_d) \Big) \leq \xi_{\text{Obj}_\text{Den}},~~~~~~~~~~~~~~~~~~~~~~~~~~~~~~~~~~~~\forall t_d\in\{1,\ldots,T_D\},
			\end{align}
		\begin{align}
			\label{stochastic_PS_OMA_C1}&\text{Pr} \Big( {B}^{}_p(t_d)-T_s\big(\beta+\frac{(1-\beta)}{2}\big)p_{pp}(t_d)\leq 0 \Big) \leq \xi_\text{C1},~~~~~~~~~~~~\forall t_d\in\{1,\ldots,T_D\},\\
			\label{probab_c4}&\text{Pr} \Big( {B}_s^{}(t_d)+{T_s}	\beta\eta_{1,s}\eta_{2,s}p_{pp}(t_d) {h}_{\text{PAP}\rightarrow \text{SAP}}(t_d)-\nonumber\\
			&~~~~~~~~~~~T_s\frac{(1-\beta)}{2}\big(p_{ss}(t_d)+p_{sp}(t_d)\big) 
			\leq 0\Big) \leq  \xi_\text{C2},~~~~~~~~~~~~~~~~~~~~~~~~~\forall t_d\in\{1,\ldots,T_D\},\\
			\label{probab_c5}&\text{Pr}\Big({B}_p^{}(t_d)  + \eta_{1,p}E^{\text{RE}}_p(t_d) -T_s\big(\beta+\frac{(1-\beta)}{2}\big)p_{pp}(t_d) \geq B_{\max,p}  \Big) \leq  \xi_\text{C3},~~~~\forall t_d\in\{1,\ldots,T_D\},\\
			\label{probab_c6}&\text{Pr}\Big({B}_s^{}(t_d)  +{T_s}\beta\eta_{1,s}\eta_{2,s}\sum\limits_{i=t_d}^{t_d+1}p_{pp}(i) {h}_{\text{PAP}\rightarrow \text{SAP}}(i)- \nonumber\\
			&~~~~~~~~~~~~~~T_s\frac{(1-\beta)}{2} \big(p_{ss}(t_d)+p_{sp}(t_d)\big)  \geq B_{\max,s}  \Big) \leq  \xi_\text{C4},~~~~~~~~~~~\forall t_d\in\{1,\ldots,T_D\},\\
			\label{propbabc7}& \text{Pr}\Big( \sum\limits_{t_d=1}^{T_D} \min\{r_{ps}^{}(t_d),r^{\text{MRC}} (t_d)\} \leq R_{\min,p} \Big) \leq  \xi_\text{C5},~~~~~~~~~~~~~~~~~~~~~~~\forall t_d\in\{1,\ldots,T_D\},
		\end{align}
	\end{subequations}
where $\xi_{[\cdot]}$ is the amount of threshold that probability constraint allow to violate. By adjusting $\xi_{[\cdot]}$, we can make the network becomes more robust against uncertainty or improve the final goal of optimization problem. {\eqref{probab_c1} and \eqref{probab_c2} are constraints related to the nominator and the denominator of \eqref{O_TS_WET_NOMA_OF} and determine the probability of violating. \eqref{stochastic_PS_OMA_C1} and \eqref{probab_c4} show the amount of probability that the causality constraints can violate. \eqref{probab_c5} and \eqref{probab_c6} relate to the violation of the batteries overflow constraints. Finally, the probability of violation related to the PUE's rate is described in  \eqref{probab_c6}. }
	
We use the chance constraint approach \cite{ben2008selected} to relax the probability nature of uncertain variables. Here, we apply this approach to one of the uncertain variable and expand it to the all of the uncertain variables in the constraints. 
	By applying this approach to \eqref{stochastic_PS_OMA_C1} we have 
	\begin{align}
		{B}^{}_p(t_d)-T_s\big(\beta+\frac{(1-\beta)}{2}\big)p_{pp}(t_d)= \bar{B}^{}_p(t_d)+\zeta_{B^{}_p}  \epsilon_{{B}^{}_p(t_d)}-T_s\big(\beta+\frac{(1-\beta)}{2}\big)p_{pp}(t_d),
	\end{align}
	where $\zeta_{B^{}_p}=\frac{{B}^{}_p(t_d)-\bar{B}^{}_p(t_d)}{\epsilon_{{B}^{}_p(t_d)}} \in [-1,1]$.  $\zeta_{B^{}_p}$ follows the probability distributed function $\mathcal{F}_{\zeta_{B^{}_p}}$.  
	By adopting Bernstein approximation, the constraint can be {reformulated} as follows:
	\begin{align}
		\bar{B}^{}_p(t_d)+
		\chi^+_{\zeta_{B^{}_p}  } {\epsilon_{{B}^{}_p(t_d)}}+\sqrt{2\text{ln}\frac{1}{\xi_\text{C1}}}\big({\tau_{\zeta_{B^{}_p}}}^2{\epsilon_{{B}^{}_p(t_d)}}^2\big)^\frac{1}{2}-T_s\big(\beta+\frac{(1-\beta)}{2}\big) \geq 0,
	\end{align}
	where $\chi^+_{\zeta_{B^{}_p}} \in [ 0 , 1 ]$ and $\tau_{\zeta_{B^{}_p}}\in [0, \infty] $ are {chosen to obtain a safe approximation} and are given in Table \ref{typical_probability_distribution}. 
	
\begin{table}[h]
\caption{Values of  $\chi^+_{\zeta_{a}}$ and $\tau_{\zeta_{a}}$ for typical probability distribution $\mathcal{F}_{\zeta_{a}}$ }
\small
\label{typical_probability_distribution}
\centering
\begin{tabular}{  l  c  c   } 
	~~~~~~~~~~~~$\mathcal{F}_{\zeta_{a}}$ & $\chi^+_{\zeta_{a}}$ &  $\tau_{\zeta_{a}}$ \\ 
	\hline
	1. $\mathcal{F}_{\zeta_{a}}\in[-1,1]$ & 1 & 0 \\ 
	\hline
	2. $\mathcal{F}_{\zeta_{a}}$ is {unimodal probability distribution and} $\mathcal{F}_{\zeta_{a}}\in[-1,1]$  & $\frac{1}{2}$ & $\frac{1}{\sqrt{12}}$  \\ 
	\hline
	3. $\mathcal{F}_{\zeta_{a}}$ is symmetric, unimodal probability distribution and $\mathcal{F}_{\zeta_{a}}\in[-1,1]$ & 0 & $\frac{1}{\sqrt{3}}$  \\ 
	\hline
\end{tabular}
\end{table}

	By adopting this approximation, the resource allocation problem can be reformulated as follows:
	\begin{subequations}\label{O_TS_WET_NOMA4}
		\begin{align}
			\max_{\boldsymbol{p}_{pp},\boldsymbol{p}_{sp},\boldsymbol{p}_{ss},\beta,\boldsymbol{u},\boldsymbol{v}}~~&\sum\limits_{t_d=1}^{T_D}\log_2\Big(\frac{u(t_d)}{v(t_d)}\Big),\\
			\text{s.t.~~}&	\text{C6, C7, C8,}\nonumber\\
			&\bar{\bar{R}}^{}(t_d) \geq u(t_d),~~~~~~~~~~~~~~~~~~~~~~~~~~~~~~~~~~~~~~~~~~~~~~~~~~\forall t_d\in\{1,\ldots,T_D\}, \\
			&\bar{\bar{E}}^{}(t_d) \leq v(t_d),~~~~~~~~~~~~~~~~~~~~~~~~~~~~~~~~~~~~~~~~~~~~~~~~~~\forall t_d\in\{1,\ldots,T_D\},\\
			&B_p^{}(t_d)+\Xi_{B_p^{}}  -T_s\big(\beta+\frac{(1-\beta)}{2}\big)p_{pp}(t_d)\geq 0,~~~~~~~~~~~~~~~\forall t_d\in\{1,\ldots,T_D\},\\
			&B_s^{}(t_d)+\Xi_{B_s^{}}+{T_s}	\eta_{1,s}\beta\eta_{2,s}p_{pp}(t_d) \big(\bar{h}_{\text{PAP}\rightarrow \text{SAP}}(t_d)+\Xi_{h_{\text{PAP}\rightarrow\text{SAP}}}\big)-\nonumber\\
			&T_s\frac{(1-\beta)}{2}\big(p_{ss}(t_d)+p_{sp}(t_d)\big)\geq 0 ,~~~~~~~~~~~~~~~~~~~~~~~~~~~~~\forall t_d\in\{1,\ldots,T_D\},\\
			& B_p^{}(t_d) +\Xi_{B_p^{}} + \eta_{1,p}E^{\text{RE}}_p(t_d) -T_s\big(\beta+\frac{(1-\beta)}{2}\big)p_{pp}(t_d) \leq B_{\max,p} ,~~\forall t_d\in\{1,\ldots,T_D\},\\
			&B_s^{}(t_d) +\Xi_{B_s^{}} +{T_s}\eta_{1,s}\beta\eta_{2,s}\sum\limits_{i=t_d}^{t_d+1}p_{pp}(i) \big(\bar{h}_{\text{PAP}\rightarrow \text{SAP}}(i)+\Xi_{h_{\text{PAP}\rightarrow\text{SAP}}}\big) \nonumber\\
			&-T_s\frac{(1-\beta)}{2} \big(p_{ss}(t_d)+p_{sp}(t_d)\big) \leq B_{\max,s}  ,~~~~~~~~~~~~~~~~~~~~\forall t_d\in\{1,\ldots,T_D\},\\
			& \sum\limits_{t_d=1}^{T_D} \bar{\bar{R}}_{\min}^{}(t_d) \geq R_{\min,p},~~~~~~~~~~~~~~~~~~~~~~~~~~~~~~~~~~~~~~~~~~\forall t_d\in\{1,\ldots,T_D\},
		\end{align}
	\end{subequations}
	where
	\begin{align}
		&\bar{\bar{R}}^{}(t_d)=\frac{(1-\beta)}{2}\Big[\bar{\bar{r}}_{\text{SAP}\rightarrow \text{SUE}} ^{\text{S}}(t_d)-\log_2\Big(1+\frac{p_{ss}(t_d)(\bar{h}_{\text{SAP}\rightarrow E}(t_d)+\Xi_{h_{\text{SAP}\rightarrow E}}) } {\sigma_{se}^2}\Big)\Big]^+ +\\
		&\frac{(1-\beta)}{2}\Big[\bar{\bar{R}}^{}_{\min}(t_d)-\log_2\Big(1+\frac{p_{pp}(t_d)(\bar{h}_{\text{PAP}\rightarrow E}(t_d)+\Xi_{h_{\text{PAP}\rightarrow E}})}{\sigma_{pe}^2}+ \frac{p_{sp}(t_d)(\bar{h}_{\text{SAP}\rightarrow E}(t_d)+\Xi_{h_{\text{SAP}\rightarrow E}}) } {\sigma_{se}^2} \Big)\Big]^+,\nonumber\\
		&\bar{\bar{E}}^{}(t_d)={T_s}\nu \Big( \big(\frac{(1-\beta)}{2}+\beta\big)p_{pp}(t_d)+\frac{(1-\beta)}{2} p_{sp}(t_d) +\frac{(1-\beta)}{2}p_{ss} (t_d) \Big) -\\
		&~~~~~~~~~~ {T_s}\nu  \beta  \eta_{2,s}  p_{pp}(t_d) \big(\bar{h}_{\text{PAP}\rightarrow\text{SAP}}(t_d)+\Xi_{h_{\text{PAP}\rightarrow\text{SAP}}}\big)  + E_c,\nonumber
	\end{align}
\begin{align}
		&\bar{\bar{R}}^{}_{\min}(t_d)=\min\Big\{\log_2\Big(1+\frac{p_{pp}(t_d)(\bar{h}_{\text{PAP}\rightarrow\text{SAP}}(t_d)+\Xi_{h_{\text{PAP}\rightarrow\text{SAP}}})}{\sigma_{ps}^2}\Big),\bar{\bar{r}}_{\text{MRC-P}}^{\text{S}} (t_d)\Big\},\\	 	
		&\bar{\bar{r}}_{\text{SAP}\rightarrow \text{SUE}} ^{\text{SCN I}}(t_d)=\log_2\Big(1+\frac{p_{ss}(t_d)\big(\bar{h}_{\text{SAP}\rightarrow \text{SUE}}(t_d)+\Xi_{h_{\text{SAP}\rightarrow \text{SUE}}}\big)}{p_{sp}(t_d)\big(\bar{h}_{\text{SAP}\rightarrow\text{SUE}}(t_d)+\Xi_{h_{\text{SAP}\rightarrow \text{SUE}}}\big)+\sigma_{ss}^2}\Big),\\
		&\bar{\bar{r}}_{\text{SAP}\rightarrow \text{SUE}} ^{\text{SCN II}}(t_d) = \log_2\Big(1+\frac{p_{ss}(t_d)\big(\bar{h}_{\text{SAP}\rightarrow \text{SUE}}(t_d)+\Xi_{h_{\text{SAP}\rightarrow \text{SUE}}}\big)}{\sigma_{ss}^2} \Big),\\
		&\bar{\bar{r}}_{\text{MRC-P}}^{\text{SCN I}}(t_d)=\log_2\big( 1+ \frac{p_{pp}(t_d)\big(\bar{h}_{\text{PAP}\rightarrow\text{PUE}}(t_d)+\Xi_{h_{\text{PAP}\rightarrow\text{PUE}}}\big)}{\sigma^2_{pp}} + \frac{p_{sp}(t_d)\big(\bar{h}_{\text{SAP}\rightarrow \text{PUE}}(t_d)+\Xi_{h_{\text{SAP}\rightarrow \text{PUE}}}\big)}{\sigma_{sp}^2} \big),
	\end{align}
\begin{align}
		&\bar{\bar{r}}_{\text{MRC-P}}^{\text{CN II}}(t_d)=\log_2\Big( 1+ \frac{p_{pp}(t_d)\big(\bar{h}_{\text{PAP}\rightarrow\text{PUE}}(t_d)+\Xi_{h_{\text{PAP}\rightarrow\text{PUE}}}\big)}{\sigma^2_{pp}} + \frac{p_{sp}(t_d)\big(\bar{h}_{\text{SAP}\rightarrow \text{PUE}}(t_d)+\Xi_{h_{\text{SAP}\rightarrow \text{PUE}}}\big)}{p_{ss}(t_d)\big(\bar{h}_{{\text{SAP}\rightarrow \text{PUE}}}(t_d)+\Xi_{h_{\text{SAP}\rightarrow \text{PUE}}}\big)+\sigma_{sp}^2} \Big).
	\end{align}
{and  $\Xi_{[a]}	 = \chi^+_{\zeta_{[a]}}   {\epsilon_{[a]}}+\sqrt{2\text{ln}\frac{1}{\xi_{[\cdot]}}}\big({\tau_{\zeta_{[a]}}}^2{\epsilon_{[a]}}^2\big)^\frac{1}{2}$ where  $\xi_{[\cdot]}, [\cdot] \in \{\text{Obj}_\text{Nom}, \text{Obj}_\text{Den} \text{C1}, \text{C2}, \text{C3}, \text{C4}, \text{C5} \}$  is violation threshold related to each constant of \eqref{Ps_OMA_stochastic},} and $[a] \in \{B_p, B_s, h_{\text{PAP}\rightarrow \text{SAP}},h_{\text{SAP}\rightarrow \text{PUE}}, h_{\text{SAP}\rightarrow \text{SUE}}, h_{\text{PAP}\rightarrow E}, h_{\text{SAP}\rightarrow E},\}$.  $\chi^+_{\zeta{[a]}}$ and $\tau_{\zeta_{[a]}}$ are given in Table \ref{typical_probability_distribution}.
	
	\section{Solution and Algorithm}\label{Solution_algorithm}
	\subsection{Background of Machine Learning }\label{Background_Machine_Learning}
	
	\subsubsection{POMDP}
	In multi-agent reinforcement learning, a group of agents cooperatively engage in an environment to reach a common goal. The agents act independently, and there is no implicit communication or information sharing between them. The agents do not know the underlying true state of the environment, but rather have individual observations correlated with the state. The objective of the agents is to reduce uncertainty about the hidden state. Such problems are appropriately modeled as decentralized partially observable Markov decision processes (Dec-POMDPs) \cite{Oliehoek2016}. 
	
	The POMDP is a Markov decision process (MDP) in which the state is partially observable to the agent and allows for choosing action under uncertain conditions. The  MDP can be described as a tuple $\big \langle \mathcal{S},\mathcal{A}, \mathcal{T}, \mathcal{R}\big \rangle$, where $\mathcal{S}$ is the state space of the environment, $\mathcal{A}$ is the set of agent actions, $\mathcal{T}$ is the state transition probability and $\mathcal{R}$ is the reward function. At each time step, the agent selects an action {$a_{t} \in \mathcal{A}$} based on the observation $o_t \in \mathcal{S}$ and receives an immediate reward.	
	%\subsection{Histories for Dec-POMDP}
	In the MDP, the agent chooses an action based on a policy $\pi$ that maps state to action\cite{Oliehoek2016}. However, in the POMDP, since the agent does not have the information of full state, it utilizes a belief state in order to predict the actions \cite{astrom1965optimal, Oliehoek2016}. The belief state encapsulates all of the information about the agent's states and actions history $h_t = (a_{0},o_{1},\dots,o_{t-1},a_{t-1},o_{t})$. A policy {$\pi$} for agent maps the histories to actions. More precisely, at  time step $t$, the probability of choosing an action {$a_{t}$} given the history of actions and observation $h_t$ until time step $t$, is defined as $\pi(a_{t}|h_t)$.
	The objective of the POMDP in state $\mathcal{S}$, given  action $\mathcal{A}$ is to find a policy that maximizes its expected cumulative reward as
	\begin{align}
		\pi^* = \argmax_{\pi}\mathbb{E}_{{\mathcal{S}_{t}},{\mathcal{A}_{t}}} \big[\sum_{t}^{\infty} \mathcal{R}(\mathcal{S}_{t},\mathcal{A}_{t}) \big].
	\end{align}
	It is impractical to precisely find a solution for the above objective, multi agent deep reinforcement learning {(MA-DRL)} methods can be used as a sub-optimal alternative {to single agent DRL solutions} \cite{lee2020multi}. Another solution is to use  the RNN instead of feed forward neural network and consider the information of the agent's state and action histories $h_t$  for action selection \cite{heess2015memory}. Since the state of the environment of our proposed model is partially observable and there are two different objectives for the agents as individual and common objectives, {we utilize the MASRDDPG which is a {derivative} {of} the decomposed multi agent deep deterministic policy gradient (DE-MADDPG)} proposed in \cite{sheikh2020multi}. For further analyzes, we consider {the} RDPG method as {a} second solution since it is suitable for partially observable and uncertain {environments} \cite{heess2015memory}.
	
	\subsubsection{MASRDDPG}
	In our proposed solution, there are two agents, the PAP and the SAP with policies {$\pi = \{ \pi_1, \pi_2 \}$} parameterized by  {$\theta = \{ \theta_1, \theta_2\}$}, that cooperate with each other in order to serve the PUE by considering the energy consumption and other constraints. The policy gradient for the PAP is as follows:
{	\begin{align}
		\nabla J(\theta_{\text{PAP}}) = \mathbb{E}_{{S,a}\sim D} \Big[ \nabla_{\theta_\text{PAP}} \pi_\text{PAP} (a_\text{PAP} |o_\text{PAP}) \nabla_{a_\text{PAP}} Q_\text{G}^{\pi} (S, a_\text{PAP}, a_\text{SAP})|_{{a_\text{PAP}}=\pi_\text{PAP} (o_\text{PAP})}\Big]+\nonumber\\
		\mathbb{E}_{{S,a}\sim D} \Big[ \nabla_{\theta_\text{PAP}} \pi_\text{PAP} (a_\text{PAP} |o_\text{PAP}) \nabla_{a_\text{PAP}}  {Q_\text{PAP}^{\pi} (o_\text{PAP}, a_\text{PAP})}|_{{a_\text{PAP}}=\pi_\text{PAP} (o_\text{PAP})}\Big],
	\end{align}}
	where $S=(o_\text{PAP}, o_\text{SAP})$ and {$Q_\text{G}^{\pi}(S, a_\text{PAP}, a_\text{SAP})$} is the centralized action-value function (critic network) parameterized by {$\omega$} which estimates the Q-value of both agents, the PAP and the SAP. $Q^{\pi}_\text{PAP}$ is the local action-value function of the PAP which is parameterized by $\omega^\text{PAP}$ that estimates the Q-value for the PAP. {$D$ represents the experience replay buffer and the expectation is taken over a random mini-batch of $D$.
	Considering a local action-value function for the SAP as $Q^{\pi}_\text{SAP}$, the policy gradient for the SAP can be written as}
{{\begin{align}
		\nabla J(\theta_\text{SAP})= \mathbb{E}_{{S,a}\sim D} \Big[ \nabla_{\theta_\text{SAP}} \pi_\text{SAP} (a_\text{SAP} |o_\text{SAP}) \nabla_{a_\text{SAP}} Q_\text{G}^{\pi} (S, a_\text{SAP}, a_\text{PAP})|_{{a_\text{SAP}}=\pi_\text{SAP} (o_\text{SAP})}\Big]\nonumber +\\
		 \mathbb{E}_{{S,a}\sim D} \Big[ \nabla_{\theta_\text{SAP}} \pi_\text{SAP} (a_\text{SAP} |o_\text{SAP}) \nabla_{a_\text{SAP}}  {Q_\text{SAP}^{\pi} (o_\text{SAP}, a_\text{SAP})}|_{{a_\text{SAP}}=\pi_\text{SAP} (o_\text{SAP})}\Big].
	\end{align}}}
	The global critic of {the PAP and the SAP} is updated based on the following equation: 
	\begin{align}
		\mathcal{L}(\omega)= \mathbb{E}_{s, a, r, s^{\prime}} \Big[(Q_{G}^{\pi}(S, a_\text{PAP}, a_\text{SAP})-y_{g})^2\Big],
	\end{align}
	where {$y_g =r_g+\gamma {Q_{G^{\prime}}^{\pi}(S^{\prime}, a_{\text{PAP}^{\prime}})|_{a_\text{SAP}^{\prime}=\pi_{\text{SAP}}^{\prime}(o_\text{SAP}^{\prime})}}$}. {$\pi^{\prime}=\{\pi_\text{PAP}, \pi_\text{SAP}\}$} {contains the} target policies parameterized by target actors with parameters {$\theta^{\prime} = {\{\theta_\text{PAP}^{\prime}, \theta_\text{SAP}^{\prime}\}}$}, $Q_{G'}^{\pi}$ is the target global critic parameterized by $\omega'$, $r_g$ is the immediate global reward and $0 <\gamma<1$ is the discount factor. Finally, the {local critic update for the SAP} is as follows:
	\begin{align}\label{local_1}
		\mathcal{L}(\theta)= \mathbb{E}_{o, a, r, o^{\prime}}\Big[\big(Q_\text{SAP}^{\pi}(o_\text{SAP}, a_\text{SAP}) - y_{l}\big)^2\Big],
	\end{align}
	where {$y_{l}=r_{l}^\text{SAP}+{\gamma}Q_{{\text{SAP}}^{\prime}}^{\pi^{\prime}}(o_\text{SAP}^{\prime}, o_\text{SAP}^{\prime})|_{a_\text{SAP}^{\prime}={\pi}^{\prime}(o_\text{SAP})^{\prime}}$}, $Q_{{\text{SAP}}^{\prime}}^{\pi^{\prime}}$ is the target critic for the SAP that is parameterized by ${\omega_{\text{SAP}}'}$ and $r_l^{\text{SAP}}$ is the local reward for the SAP.
	
	The goal of this combination is to maximize the shared and individual rewards, i.e., shared reward for serving PUE and the individual reward {for the SAP to serve the SUE.} Similarly, the local critic update for the PAP is calculated based on (\ref{local_1}).
	
	{\subsubsection{RDPG}
	As mentioned above, the POMDP prescribes the optimal action for each transition observation based on  a policy $\pi(h_t)$, { $h_t = (o_1, a_1, ..., s_{t-1}, o_{t-1}, a_{t-1})$}, and the histories of observations in order to maximize the discounted expected cumulative reward as follows:}   
	%The solution of the POMDP is a policy prescribing which action is optimal for each transition history on a policy $\pi(h_t)$ which maps the histories of observations into action \cite{heess2015memory}, in order to maximize the discounted cumulative reward as follows:
	\begin{align}
		J = \mathbb{E}_{\tau}\Big[\sum_{t=1}^{\infty} \gamma^{t-1} r(s_t, a_t)\Big],
	\end{align}
	where $\tau$ is the trajectory of all transitions, $(s_1,o_1, a_1, s_2, o_2, a_2, ..., s_{\mathfrak{L}}, o_{\mathfrak{L}}, a_{\mathfrak{L}})$, $\mathfrak{L}$ is the trajectory length which can be defined by experiments, and the expectation is taken over the trajectory.  
	%$h_t = (o_1, a_1, ..., s_{t-1}, o_{t-1}, a_{t-1})$ is the entire history till time step t.
	The trajectory $\tau$ is obtained from the trajectory distribution influenced by the policy $\pi$ as $p(s_1)p(o_1|s_1)\pi(a_1|h_1)p(s_2|s_1,a_1)\\p(o_2|s_2)\pi(a_2|h_2)\dots$~. The action-value function in the POMDP can be written as follows:
	\begin{align}
		Q^{\pi}(h_t, a_t) = \mathbb{E}_{s_t|h_t}[r_t(s_t, a_t)]+\mathbb{E}_{\vartheta|h_t,a_t}\Big[\sum_{i=1}^{\mathfrak{L}} \gamma^{i} r(s_{t+i}, a_{t+i})\Big],
	\end{align}
	where $\vartheta = (s_{t+1}, o_{t+1}, a_{t+1}, ...)$ is the future trajectory after $t$. 
	
	The deterministic policy gradient (DPG) with the RNN can be extended to handle the environment with partially observability and uncertainty \cite{heess2015memory}. The key of using the {RNN} instead of
	%feed forward one in order to provide the ability for the agent to learn from the past history in POMDP environment.
	{feed forward method is to provide the ability of learning form the past history. } The policy for action selection can be updated as follows:
	\begin{align}\label{equation-policy-update}
		\nabla_{\theta}J(\theta)=\mathbb{E}_{\tau}[\sum_{t=1}^{\infty} \gamma^{t-1} \nabla_{a}Q^{\mu}(h_{t}, a)|_{a=\mu^{\theta}(h_t)} \nabla_{\theta} \mu^{\theta}(h_t)].
	\end{align}
	For the stability and efficiency, the experience replay is used for the  {DPG} {method. During} the learning process the experienced histories are stored in a replay buffer then the expectat{ion i}n  (\ref{equation-policy-update}) is taken over a sampled 
	{trajectory experiences} \cite{heess2015memory, hafner2011reinforcement}. 
	Since the {DPG } method has actor (policy $\pi$) and critic (value function $Q$) networks with parameter $\theta^{\mu}$ and $\theta^{Q}$, respectively, %therefore 
	there are two copies of actor and critic with parameters $\theta^{\mu^{\prime}}$ and $\theta^{Q^{\prime}}$, respectively \cite{hafner2011reinforcement}. The target actor network is updated using back propagation through time over $N$ sampled episodes with duration $T$ as follows:
	{\begin{align}
			\nabla{\theta^{\mu}} = \frac{1}{NT_d} \sum_{i=1}^{N} \sum_{t_d=1}^{T_d} \nabla_{a} Q^{\theta^Q}(h_{t_d}^i, \mu^{\theta^{\mu}}(h_{t_d}^{i})) \nabla _{\theta^{\mu}} \mu^{\theta^{\mu}}(h_{t_d}^{i}).
	\end{align}}
	Similarly, the critic target network is updated using following equation:
	{\begin{align}
			\nabla{ \theta^{Q}} = \frac{1}{NT_d} \sum_{i=1}^{N} \sum_{t_d=1}^{T_d} (y_{t_d}^{i}-Q^{\theta^Q}(h_{t_d}^{i}, a_{t_d}^{i})) \nabla \theta^Q Q^{\theta^Q}(h_{t_d}^i, a_{t_d}^i),
	\end{align}}
	{where {$y_{t}^{i} = r_{t}^i+\gamma Q^{\theta^{ Q^{\prime}}}(h_{t+1}^i, \mu^{\theta^{\prime}}(h_{t+1}^i))$} are the target values for each sample $i$, $r_t^i$. is the immediate reward at time slot $t$ for $i^{th}$ sample, $Q^{\theta^{Q'}}$ denotes the target critic parameterized by $\theta^{Q'}$, and $\mu^{\theta'}$ is the deterministic policy for target actor network parameterized by $\theta'$. }
	\subsection{Problem environment}
	{We consider the energy harvested secure cooperative downlink transmission as the POMDP in which the PAP and the SAP play the agents role in the environment which consists of one PUE, one SUE and one eavesdropper with their channel gains and data rate requirements. The PAP and the SAP are {described by the state consisting} of the PUE, the SUE, and the eavesdropper channel gains, their data rate requirements and self battery energy values. Based on their observations, they select their own action separately at each time slot $t$ and receive a shared reward $r_g(t)$ along with their individual reward $r_l(t)$. Note that the eavesdropper channel gain {are} not observable neither {by} the PAP nor the SAP. The state space, action space, and reward function of our MASRDDPG method are explained as follows:
		\begin{itemize}
			\item \textbf{State:}  The state of the PAP and the SAP at time slot $t$ can be shown as $s^{(t)}_{\text{PAP}} \in \mathcal{S}$, $s^{(t)}_\text{SAP}\in \mathcal{S}$. The state of each agent is the observation of that agent from the environment. The state space is the set of states that {is} {observed} from the environment by the agents and consist of the channel gains of transmitters to receivers and the eavesdroppers, the amount of battery energy at the PAP and the SAP, and the amount of harvested energy at the PAP. The state space of the PAP and the SAP {are} as follows:
			\begin{align}
				s^{(t)}_{\text{PAP}}&=\{{\bar{h}}_{\text{PAP}\rightarrow \text{PUE}}(t_d),{\bar{h}}_{\text{PAP}\rightarrow \text{SAP}}(t_d),{\bar{B}}_{p}(t_d)\},\\
				s^{(t)}_{\text{SAP}}&=\{{\bar{h}}_{\text{SAP}\rightarrow \text{PUE}}(t_d),{\bar{h}}_{\text{SAP}\rightarrow \text{SUE}}(t_d),{\bar{B}}_{{s}}(t_d)\}.
			\end{align}
		Note that the agents observation of channel gains and battery energy values suffer from uncertainty. 
			\item \textbf{Action}: Each agent has its own action. The action of the PAP and the SAP at time slot $t$ {are}
			\begin{align}
				a_\text{PAP}^{(t)}&=\{p_{pp}(t_d),\beta\},\\
				a_\text{SAP}^{(t)}&=\{p_{sp}(t_d),p_{ss}(t_d)\}.
			\end{align}
			\item[] The action space is a set of actions that each agent selects from the space and consists of transmit power of the PAP and the SAP, as well as, the TS factor $\beta$. Here, all variables are continuous value where the power allocation variables take place in  {$[0,P^{\max}_{[\star]}], ~[\star]\in\{p,s\}$} and the range of $\beta$ factors is $[0,1]$.   
			\item \textbf{Reward}: The objective of resource allocation problems are maximizing the {PFEE} of secure transmission. Hence, the reward should be defined in a way that achieves maximum PFEE. There is a global reward $r_{\text{G}}^{(t)}$ that is the objective of the optimization problem. 
			{On the other hand, we use a learning method where agents have their individual objectives. In other words, the PAP and the SAP will be rewarded (or punished) by $r_{\text{PAP}}^{(t)}$ and $r_{\text{SAP}}^{(t)}$ as they satisfy or violate their corresponding constraints. In addition, the SAP should serve the SUE, therefore the SAP will obtain additional reward based on need to define an extra individual objective for the SAP, in order to consider the SUE requirements as well. Therefore, the total individual (local) reward for the SAP is calculated as follows:
			\begin{align}
%				r^{(t)}_{\text{G}}&=\frac{\gamma_{pp}+\frac{(1-\beta)}{T_s}\gamma_{ps}-\gamma_{\text{PAP}-E}}{p_{pp}-p_{s}}, \\
				r^{(t)}_{\text{SAP,Tot}}&=\frac{r_{\text{SAP}\rightarrow \text{SUE}} ^{\text{S}}(t)-r_{\text{SAP2}\rightarrow E }(t)}{\frac{(1-\beta)}{2}\big(p_{ss}(t)\big)}+r^{(t)}_{\text{SAP}},
			\end{align}
			\item [] and the global reward for each uncertainty approach is the objective function of the related optimization problem, i.e., (\ref{O_TS_WET_NOMA}a), (\ref{O_TS_WET_NOMA2}a), (\ref{Ps_OMA_stochastic}a), (\ref{O_TS_WET_NOMA4}a). It is worth mentioning that the local reward for the PAP is just punishment for violating its corresponding constraints such as depleting the energy of battery.}
		\end{itemize}
		}

		% \item Action: Each agent has its own action. The action of PAP and SAP for PS and TS can be shown as
		% \item[] \begin{align}
		% 	a_{PAP}^{PS}&=\{p_{pp}(t_d)\}\\
		% 	a_{SAP}^{PS}&=\{p_{sp}(t_d),p_{ss}(t_d),\alpha\}\\
		% 	a_{PAP}^{TS}&=\{p_{pp}(t_d),\beta\}\\
		% 	a_{SAP}^{TS}&=\{p_{sp}(t_d),p_{ss}(t_d)\}
		% \end{align}

		%
		%
		%The parameter of critic, $\theta^Q$, is updated by minimizing the loss function. The lost function is illustrated as 
		%\begin{align}
		%	L=\mathbb{E} \Big(y-Q(s^t, a^t)\Big)^2
		%\end{align}
		%where 
		%\begin{align}
		%	y=
		%\end{align}
		%
		%The parameter of action network, $\theta^{\mu} $ , is update by using the gradient descent method as 
		%\begin{align}
		%	\triangledown _{\theta^{\pi}} J=\mathbb{E}\Big(\triangledown_{\theta^{\mu}}\mu(o,\theta^{\mu})\triangledown_a Q(s,a,\theta^{Q}) ) \Big)
		%\end{align} 
		%
		%
		%The characteristics of this method can be explained as follows
		%\begin{itemize}
		%\item This method is the hierarchical centralized-distrusted optimum-seeking method 	which based on centralized training and decentralized execution. In this method, the critics of agents can access to actions of other agents. Moreover, the decision of actors network  base on local observation during the testing period. 
		%\item  The agents are classified from the low level agent to the high level agent. In addition, the convergence order in this algorithm is from the lowest level agent to the highest agent level.  
		%\item   The training period of this algorithm is based on the centralized training, hence, it seems that it can obtain global optimum.     
		%\end{itemize}
		Accordingly, the state space and action space of our RDPG method are defined as follows:
		\begin{align}
			&s_{\text{RDPG}}^{(t)}=\{{\bar{h}}_{\text{PAP}\rightarrow \text{PUE}}(t_d),{\bar{h}}_{\text{PAP}\rightarrow \text{SAP}}(t_d),{\bar{B}}_{p}(t_d),{\bar{h}}_{\text{SAP}\rightarrow \text{PUE}}(t_d),{\bar{h}}_{\text{SAP}\rightarrow \text{SUE}}(t_d),{\bar{B}}_{s}(t_d)\},\\
			&~~~~~~~~~~~~~~~~~~~~~~~~~~~~~~~~~~~a_{\text{RDPG}}^{ (t)}=\{p_{pp}(t_d),\beta,p_{sp}(t_d),p_{ss}(t_d)\}.
		\end{align}
		The reward function of the RDPG method is the same as global reward $r_G^{(t)}$ in the MASRDDPG method. {It is worth mentioning that the uncertain values, i.e., battery energy level and channel gain, is just in the state space of the agents and we consider the exact value to compute the rewards since the agents are in training mode.}  
		The framework of both algorithms is depicted in Algorithm.~\ref{MADDPG_algorithm} and Algorithm.~\ref{RDPG_algorithm}.
%		 Fig.~\ref{Fig:Digram_MADDGP} depicts the workflow of MASRDDPG method.
		
%		\begin{figure*}[tb]
%			\begin{center}
%				\includegraphics[width=12 cm]{Digram_MADDGP.pdf} %\vspace{-1cm}
%				\caption{Block diagram of MASRDDGP framework.}\label{Fig:Digram_MADDGP}%\vspace{-.5cm}
%			\end{center}%\vspace{-0.7cm}
%		\end{figure*}
		
{\begin{algorithm}[t]	
\caption{The MASRDDPG algorithm for resource allocation problem}
\small
\label{MADDPG_algorithm}
\textbf{Input}: Initialize weights of actors, local critics, and global critic networks, $\theta^\text{PAP}$, $\theta^\text{SAP}$, $\omega^\text{PAP}$, $\omega^\text{SAP}$ and $\omega$, with random values.\\
\textbf{Input}: Initialize weights of target actors, target local critics, and target global critic networks, $\theta^\text{PAP'}$, $\theta^\text{SAP'}$, $\omega^\text{PAP'}$, $\omega^\text{SAP'}$ and $\omega'$, with random values.\\
\textbf{Input}: Initialize the replay buffer $D$.\\
\textbf{Set}: Set $E$ as the maximum number of episodes, set $E_L$ as episode length, and $\mathfrak{E}$ as agent local network update frequency.\\
\For{ episode = 1:$E$ }{
				\For{$t$ = 1:$E_L$}{
					Receive $s^{(t)}$ of environment. \\
					Each agent $i$ receives observation $o_i^{(t)}$ from the environment.\\
					Each agent $i$ selects  $a_{i}^{(t)}=\mu(o_i^{(t)},{\theta}^{{\mu}_i})+N^{t}$ and excecutes it.\\
					Each agent $i$ receives local reward $r^{(t)}_i$ related to agent $i$. \\
					Each agent $i$ receives the new observation {$o_i^{(t+1)}$} of environment and receives global reward.\\
					Store $(o^{(t)}_\text{PAP},o^{(t)}_\text{SAP},a^{(t)}_\text{PAP},a^{(t)}_\text{SAP},r^{(t)}_\text{PAP},r^{(t)}_\text{SAP,Tot},r^{(t)}_\text{G},o^{(t+1)}_\text{PAP},o^{(t+1)}_\text{SAP})$ as an experience in experience replay $D$.
				}
Randomly sample a mini-batch from the experience replay $D$.\\
Update the global critic by minimizing\\
$L_{\text{G}}({\omega})=\mathbb{E} ((y^{(t)}_{\text{G}}-Q^{\pi}_{\text{G}}(\boldsymbol{o}^{(t)},\boldsymbol{a}^{(t)},{\omega}))^2)$ where \\
$y^{(t)}_{\text{G}}=r_{\text{G}}^{(t)}+\gamma Q^{\pi}_{\text{G'}}({\boldsymbol{o}^{\prime}}^{(t)},{\boldsymbol{a}^{\prime}}^{(t)},{\omega})$.\\
Update weight of target global critic network  by\\ {$~~~~~~~~~~~~~~~{\hat\omega}\leftarrow \tau {\omega}+ (1-\tau){\hat\omega} $}.\\
\If{episode mode $\mathfrak{E}$ then} {
\For{$i \in \{ \text{PAP}, \text{SAP}\}$}{Randomly sample a mini-batch from the replay memory $D$.\\
Update the local main critic network using Adam optimizer \cite{kingma2014adam}.\\%weight by using \eqref{loss_function}	\\
Update the local main actor network using Adam optimizer \cite{kingma2014adam}. %weight by using \eqref{gradient_actor_network}
Update the wights of target networks in actor and critic networks by \\
$\hat\theta^i \leftarrow \tau \theta^i + (1-\tau) \hat\theta^i$, and \\
$\hat\omega^{i}\leftarrow\tau \omega^{i}+(1-\tau) \hat\omega^{i} $.	
		}
			}
		}
		\end{algorithm}}
		
		\begin{algorithm}
			\caption{The  RDPG algorithm for resource allocation problem}
			\small
			\label{RDPG_algorithm}
			\textbf{Input}: Initialize weights of actor and critic networks,  $\theta^{\mu}$ and $\theta^{Q}$, with random value. \\
			\textbf{Input}: Initialize target network weights of actor and critic networks,  $\theta^{\mu^{\prime}}$ and $\theta^{Q^{\prime}}$, with random value. \\ 
			\textbf{Input}: Initialize the replay buffer $D$. \\
			\For{ e = 1:E }{
				Initialize empty history $h_0$
				\For{$t_d$ = 1:$T_d$}{
					receive observation $o_t$.\\
					$h_{t_d}\leftarrow h_{t_d-1}, a_{t_d-1}, o_{t_d}$ (store observation and previous action to the history).\\
					$a_{t_d}=\mu^{\theta}(h_{t_d})+\mathcal{N}$ (select action based on the history at time slot $t_d$ and add noise $\mathcal{N}$ for more exploration).\\			
				}
				Store the history $(o_1, a_1, r_1,..., o_{T_d}, a_{T_d}, r_{T_d})$ in D.\\
				Sample a minibatch of $\mathcal{M}$ episodes $(o_1^{i}, a_1^{i}, r_1^{i},...,{o_{T_d}}^{i}, a_{T_d}^{i}, r_{T_d}^{i})i=1,...,\mathcal{M}$ from D.		\\
				Construct histories $h_{t_d}^{i} = (o_1^i, a_1^i,..., a_{t-1}^{i}, o_{t}^{i})$.\\
				Compute target values $(y_1^{i},...,y_{T_d}^{i})$ for each sample episode using the recurrent target networks\\
				$~~~~~~~~~~~~~~~y_{t}^{i} = r_{t}^i+\gamma Q^{\theta^{ Q\prime}}(h_{t+1}^i, \mu^{\theta^{\prime}}(h_{t+1}^i))$.\\
				Update actor and critic using Adam optimizer \cite{kingma2014adam}.\\
				Update the actor and critic target networks with period $\tau$\\
				$~~~~~~~~~~~~~~~\omega^{\prime}\leftarrow \tau \omega + (1 - \tau)\omega^{\prime}$, and \\
				$~~~~~~~~~~~~~~~\theta^{\prime}\leftarrow \tau \theta + (1 - \tau)\theta^{\prime}$.\\
				
			}
		\end{algorithm}

		\subsection{Computational complexity analysis}
		In this section, we determine the computational complexity of our proposed algorithm which consists of action selection and the	training process complexity.
		\subsubsection{Computational Complexity of Action Selection}	
		{For the  MASRDDPG, there two actors, two local critics, and one global critic neural networks. We consider that each neural network is fully connected neural network with fixed number of
		$M$ hidden layers and fixed number of $N$ neurons in each hidden layer. The computational complexity of calculating the output of such neural network for an input is equal to the sum of the production of the sizes of two consecutive layers \cite{sipper1993serial}. For each agent (the PAP or the SAP), the production of the sizes of every two consecutive layers of actor and critic are $\underbrace{|S_i|\times N}_{layer1}, \dots, \underbrace{N^2}_{layern},\dots, \underbrace{N\times|A|_i}_{layerN}$ and $ \underbrace{(|S_i|+|A_i|)\times N}_{layer1}, \dots, \underbrace{N^2}_{layern},\dots, \underbrace{N\times1}_{layerN}$, respectively, where $|S|_i$ and $|A|_i$ are the size of $i^{th}$ agent's state and action spaces. Therefore,
		the computational complexity of action selection for the MASRDDPG equals the computational complexity of action selection and is given by $\mathcal{O} \big(N^2\big)$. Similarly, for RDPG method, there are one actor and one critic neural networks. Since we consider $\mathfrak{L}$ previous trajectories for action selection, the production of the sizes of every two consecutive layers of actor and critic are $\underbrace{\mathfrak{L} \times (|S|+|A|)\times N}_{layer1}, \dots, \underbrace{N^2}_{layern}, \dots, \underbrace{N \times |A|}_{layerN}$ and $\underbrace{\mathfrak{L} \times (|S|+2\times|A|)\times N}_{layer1}, \dots, \underbrace{N^2}_{layern}, \dots, \underbrace{N \times |A|}_{layerN}$, respectively, where $|S|$ and $|A|$ are the total state and action spaces, respectively. Therefore, the computational complexity of action selection for RDPG is $\mathcal{O}\big(N^2\big)$.}
		\subsubsection{Computational Complexity of Training}
	{	For a fully connected neural network, the back-propagation complexity is related to the multiplication of the input, output, and hidden layers. Based on the state and actions for the MASRDDPG method in previous section, the training process complexity is $\mathcal{O}\Big(b(|S|+|A|)N^{M}|A| \Big)$ since we consider $M$ hidden layers with $N$ neurons, and $b$ is the size of training batch. For RDPG  method, the training process complexity is $\mathcal{O}\Big(b \mathfrak{L}\times(|S|+2\times|A|)N^{M} \Big)$. }
		
		\section{Simulation {Results}}\label{Simulation_result}
		
{In this section, the simulation {setup} and results of proposed algorithm are presented\footnote{The full implementation of simulation is provided on: https://ieee-dataport.org/documents/ai-based-secure-noma-and-cognitive-radio-enabled-green-communications-channel-state-0}. }
		
		\subsection{Environment settings}
		We {choose} the network parameters for all scenarios as follows. We consider that the PAP is located at the origin (0m,0m) and the location of all others nodes are change in the network.   The channel gain is set based on the path loss model, namely $h_{\text{TRA}\rightarrow \text{REC}}d_{\text{TRA}\rightarrow \text{REC}}^{-\lambda}$  where $h_{\text{TRA}\rightarrow \text{REC}}$ is the small scale fading between the transmitter TRC and the receiver REC, $d_{\text{TRC}\rightarrow \text{REC}}$ is the distance from the transmitter TRC to receiver REC, and $\lambda$ is path-loss exponential  exponent. The renewable energy arrival at the PAP follows the uniform distribution with value $\{0 , 1U, 2U \}$, where $U$ is the unit of energy with constant value. The other simulation parameters are summarized in Table \ref{table_simulation_parameter}.

		\begin{table}[tp]
			\caption{System parameters}\label{table_simulation_parameter} \centering
			\small
			\begin{tabular}{| p{2.5cm} | p{8.5cm}| p{4.5cm} |}
				\hline
				\textbf{Parameter} & \textbf{Description} & \textbf{value} \\ \hline 
				$B$ & 	Bandwidth &	200 KHz\\ \hline
				$P_{p}^{\max}$ &  Maximum BS transmit power &	3 {W} (36 dBm) \\ \hline
				$E_c$ & The Cumulative Constant energy	& 0.1 {J} \\ \hline
				$T_s$ & Time Slot Duration &	1 ms \\ \hline
				$ \eta_{2,s}$  & Energy Conversion Efficiency for the SAP from Harvesting &	0.5 \\ \hline
				$ \eta_{2,p}$  & Energy Conversion Efficiency for the PAP &	0.6 \\ \hline
				$ \eta_{1,p}$ and $ \eta_{1,s}$  &Energy Conversion Efficiency for the PAP and the SAP From Battery to Power & 0.5 \\ \hline
				$\sigma_{pp}^2$, $\sigma_{pe}$, $\sigma_{ps}^2$, $\sigma_{sp}^2$, $\sigma_{ss}^2$ and  $\sigma_{se}^2$ & Noise Power Density&	-170 dBm/Hz\\ \hline
				$\nu$ & Amplifier efficiency	&0.9 \\ \hline
				$\lambda$ &Path Loss Exponent&	3.5\\ \hline
				$B_{p,0}$, $B_{p,0}$ & Initial Battery for the PAP and the SAP	Uniform between & uniform between (4, 20) {J}\\ \hline		
				$d_{E,P}$&Distance form the eavesdropper to the PAP& 160 m\\ \hline
				$d_{E,S}$&Distance form the eavesdropper to the SAP& 200 m\\ \hline
				$d_{p,P}$&Distance form the PUE to the PAP& 80 {m}\\ \hline
				$d_{p,S}$&Distance form the PUE to the SAP& 25 {m}\\ \hline
				$d_{s,S}$&Distance form the SUE to the SAP& 25 {m}\\ \hline
				$d_{S,P}$&Distance form the SAP to the PAP& 50 {m}\\ \hline
				
				$N$&The number of hidden layers& 3\\ \hline
				$M$&The number of neurons in each hidden layer& 512\\ \hline
				$-$&The activation function in hidden layers& ReLU\\ \hline
				$-$&The activation function in output layer& tanh\\ \hline
				$D$&The buffer size& ReLU\\ \hline
				$b$&The batch size& 64\\ \hline
				$-$&The number of episodes& 100\\ \hline
				$-$&The number of training steps (time slots)& 200000\\ \hline
				$\tau$&The target network update period& 0.001\\ \hline
			\end{tabular}%\vspace{-.5cm}
		\end{table}

		%The feasibility region of problem is restricted by these constraints. Hence, the predefined constants, $B_{\max,p}$, $B_{\max,s}$, $R_{\min,p}$ and etc. must be chosen such that there is intersection between sets.
		
		The number of  episodes and time slot is set to 100 and 200000. We assume that the environment is reset when the time slot {reaches} to its maximum value or the energy at the batteries fall below zero.
		
		%	The actor and critic neural networks for each agent are implemented based on the python Tensorflow. We design our critic network by using \textcolor{green}{...} layers which are fully connected neural network in which the number of hidden layer is \textcolor{green}{...}. The layers from input to output respectively consist of \textcolor{green}{...}, \textcolor{green}{...}, and \textcolor{green}{...} neurons.  The activation function in the input, the hidden, and output layers are \textcolor{green}{...}, \textcolor{green}{...} and  \textcolor{green}{...}, respectively.  
		\subsection{Simulation Results}
\begin{figure}[h]
			\begin{center}
				\includegraphics[width=15cm]{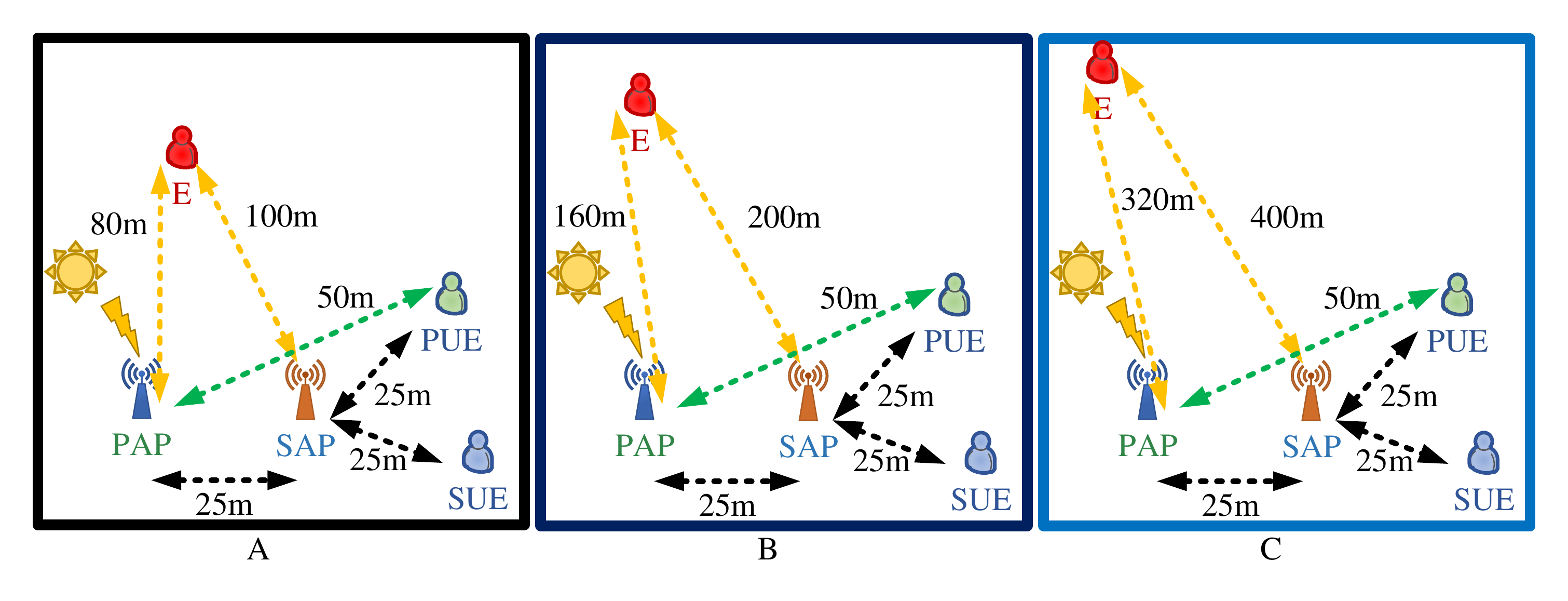}
				\caption{The simulation scenario considering the eavesdropper positions as A, B, and C.}\label{simulation_scenario}
			\end{center}
		\end{figure}

		In this section the simulation results are {discussed}. We consider the simulation scenarios as depicted in Fig.~\ref{simulation_scenario}. As {shown}, three different scenarios are considered as A, B, and C. In each scenario, the eavesdropper position changes in order to analyze the impact of its distance on the secrecy rate. It is worth mentioning that all simulation results are based on scenario B in Fig.~\ref{simulation_scenario}, except the ones which is versus the eavesdropper position.
		
%		The conservancy of RDPG, MASRDDPG, MADDPG, and DDPG methods are depicted in Fig.~\ref{PS-OMA-episodic}. The maximum transmission power of PAP and SAP are 3 watts. The amount of uncertainty is 10 percent. It can be seen that the reward related to RDPG and MASRDDPG are better than two other methods which proves the effectiveness of these two algorithms. However, the RDPG converged to a higher value of reward but with more training steps. This is because it uses the memory for predicting the actions and hence it causes more delay for its convergence, but by using the histories of the previous state transitions, i.e., state, actions and rewards, the RDPG method can learn more to the uncertainty of the environment \cite{heess2015memory}. On the other hand, the MASRDDPG converged faster than all methods since it separately trains the agents by their individual and common objectives \cite{sheikh2020multi}.
				
		Fig.~\ref{secrecy_total} demonstrates the average secrecy rate for different DRL methods considering the worst case, the stochastic, and the Bernstein approximation uncertainty models. For the sake of transparency in all figures, we consider methods with the worst case uncertainty model as "\_W" prefix, the stochastic uncertainty model as "\_S" prefix, and the Bernstein approximation uncertainty model as "\_B" prefix for all DRL methods. As depicted, all methods with the Bernstein approximation obtain higher secrecy rate compared to the stochastic and the worst case uncertainty models. By increasing the uncertainty value for all uncertainty models, all DRL methods obtain lower average secrecy rate. For example, for the uncertainty value equal to 10\% in the worst case uncertainty, MASRDDPG\_W has increased the average secrecy rate by 3.7\%, 6.7\%, and 13.1\% compared to RDPG\_W, MADDPG\_W, and DDPG\_W, respectively. On the other hand, {as the distance of the eavesdropper increases from the PAP and the SAP}, the average secrecy rate increases which is expected since the channel gain for the eavesdropper decreases remarkably. For instance,  for MASRDDPG\_S DRL method in the stochastic uncertainty model, {as the eavesdropper distance increases by two times from} the PAP and the SAP, i.e., from position A to position B, the average secrecy rate increases by 32\%. The impact of the maximum battery level on the average secrecy rate is depicted in Fig.~\ref{secrecy_total} (c). As demonstrated increasing the maximum battery level, permits the PAP and the SAP to allocate more power for the PUE and the SUE, which results in increasing the average secrecy rate for all DRL methods in all three uncertainty models. Although increasing the power will increase the eavesdropper data rate as well, the agents, i.e., the PAP and the SAP, are meant to increase the long term accumulated secrecy rate. Therefore these results are expected.
{		\begin{figure}
			\begin{center} 
				\subfigure[]{\label{sec_w}\includegraphics[width=4.5cm]{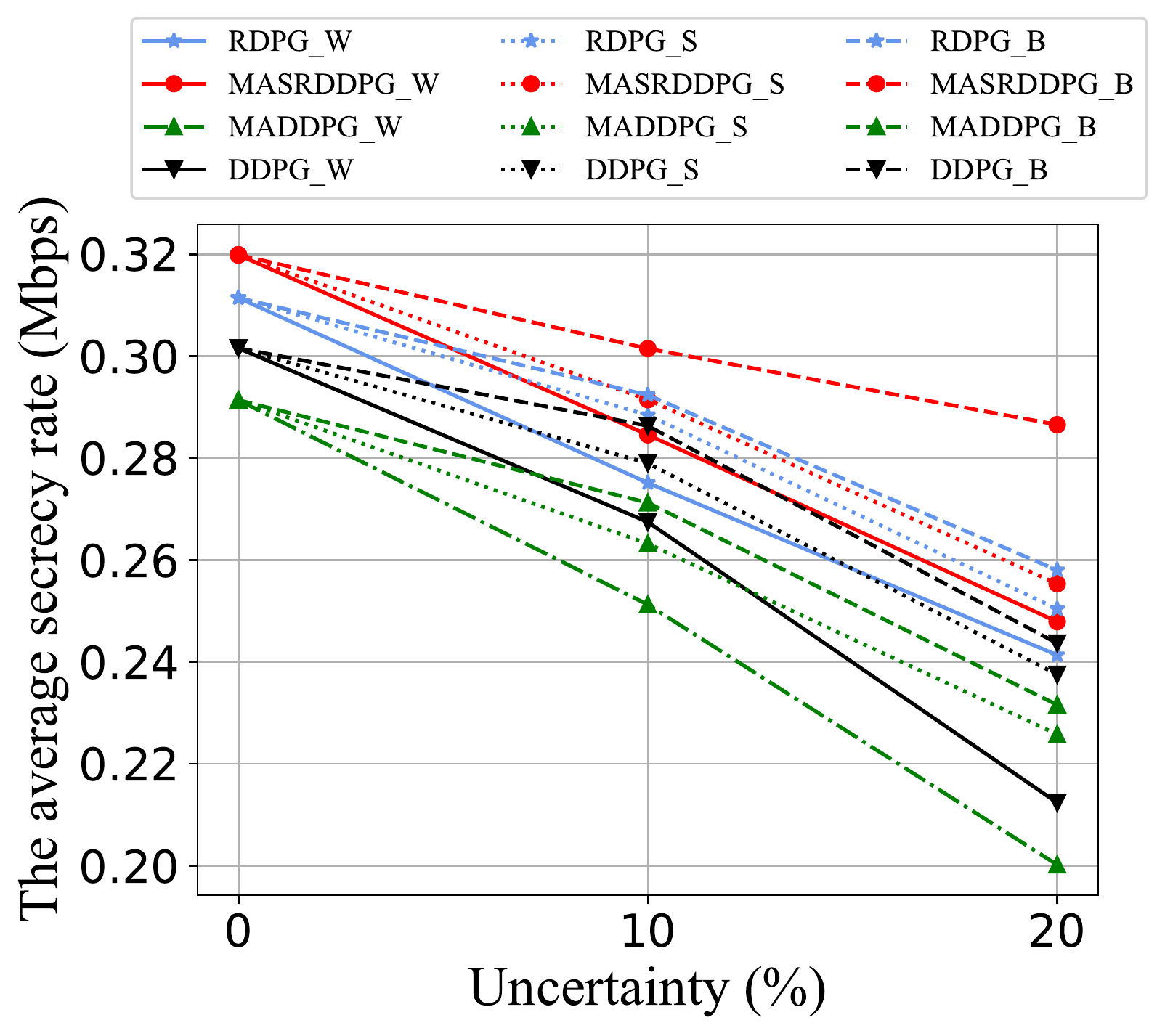}}
				\subfigure[]{\label{sec_s}\includegraphics[width=4.5cm]{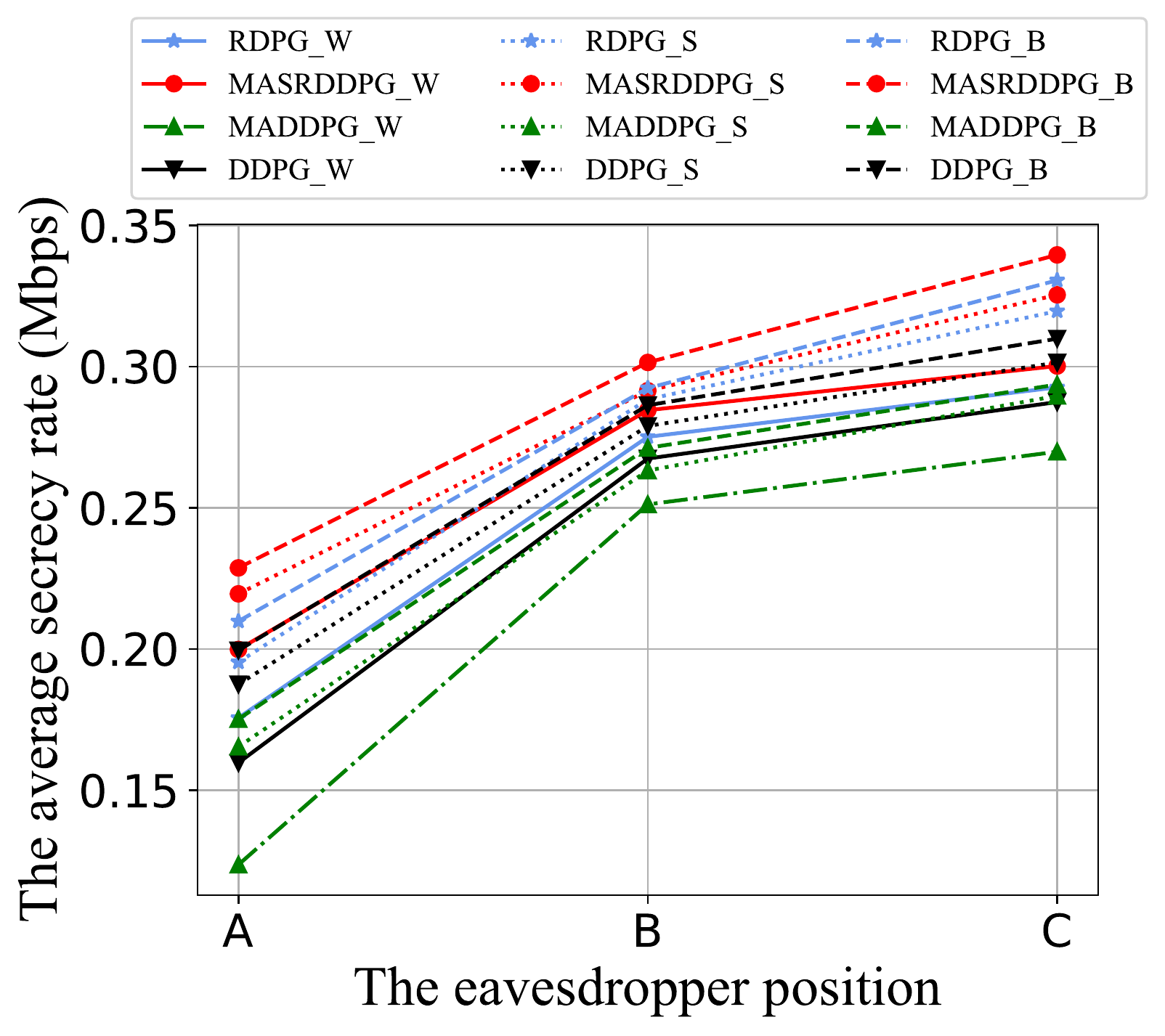}}
				\subfigure[]{\label{sec_b}\includegraphics[width=4.5cm]{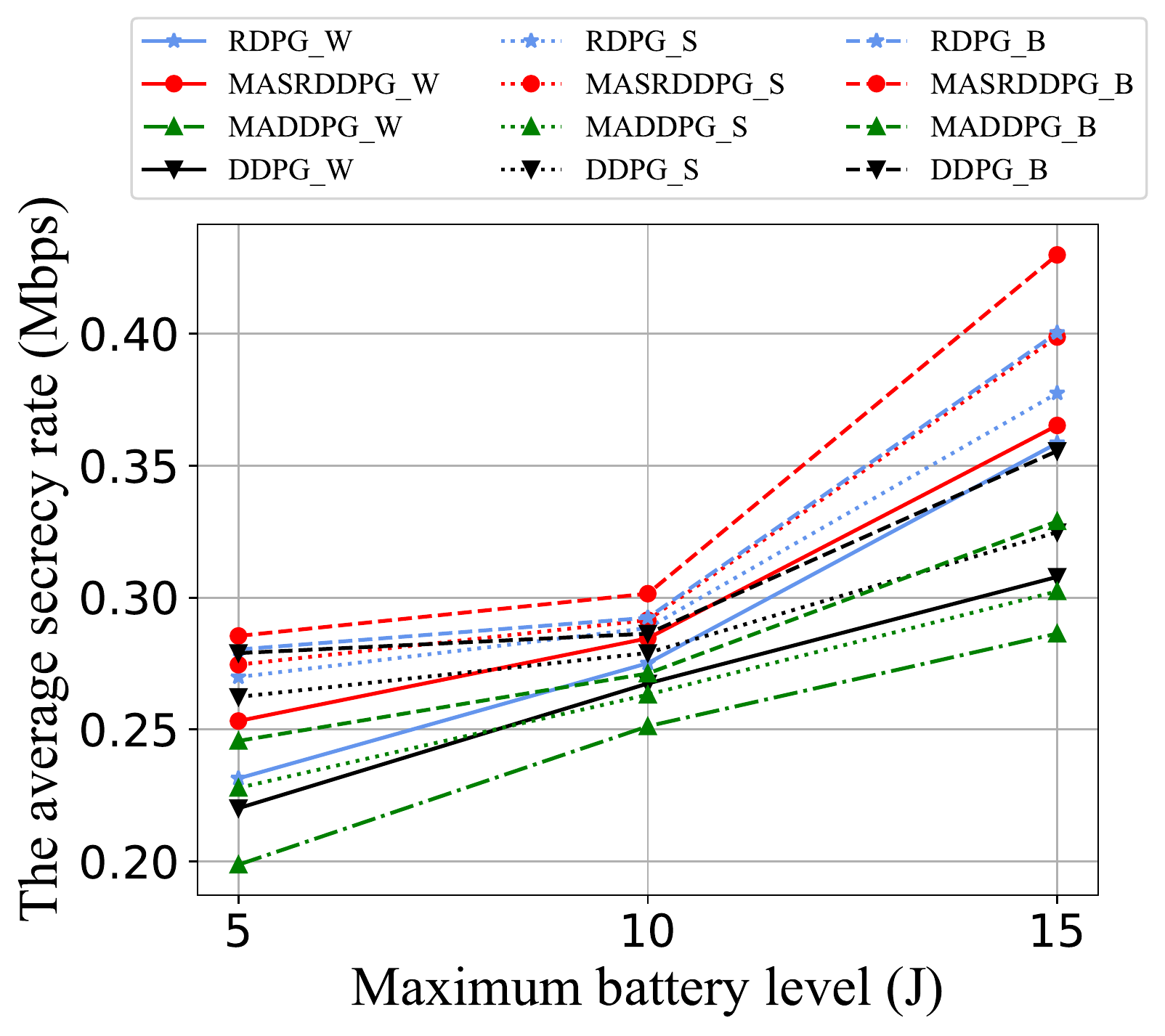}}
				\caption{The average secrecy rate for different methods with worst case "\_W", stochastic "\_S", and Bernstein "\_B" uncertainties. (a) depicts the average secrecy rate versus uncertainty value, (b) depicts versus the eavesdropper position, and (c) depicts versus maximum battery level.} \label{secrecy_total}
			\end{center}
		\end{figure}}

	The total energy consumption for all  the DRL methods in all three uncertainty models are demonstrated in Fig.~\ref{energy_total}. As demonstrated, when there is no uncertainty, all DRL methods for the specific uncertainty model, i.e., the worst case, the stochastic, and Bernstein approximation methods, have similar performance, but as the uncertainty increases, the energy consumption in the worst case model becomes higher compared to the stochastic and the Bernstein approximation model for all corresponding DRL methods, for instance, for 20\% uncertainty, the energy consumption obtained by RDPG\_W is 8.7\% and 12.1\% higher compared to RDPG\_S and RDPG\_B, respectively. Another result from the Fig.~\ref{energy_total} is, as the eavesdropper distance increases, the energy consumption decreases which is expected. On the other hand, increasing the maximum battery level allows the PAP and the SAP to consume more energy to satisfy the network and users requirements, and to increase the average secrecy rate. 
	
\begin{figure}
	\begin{center} 
		\subfigure[]{\label{sec_wd}\includegraphics[width=5.5cm]{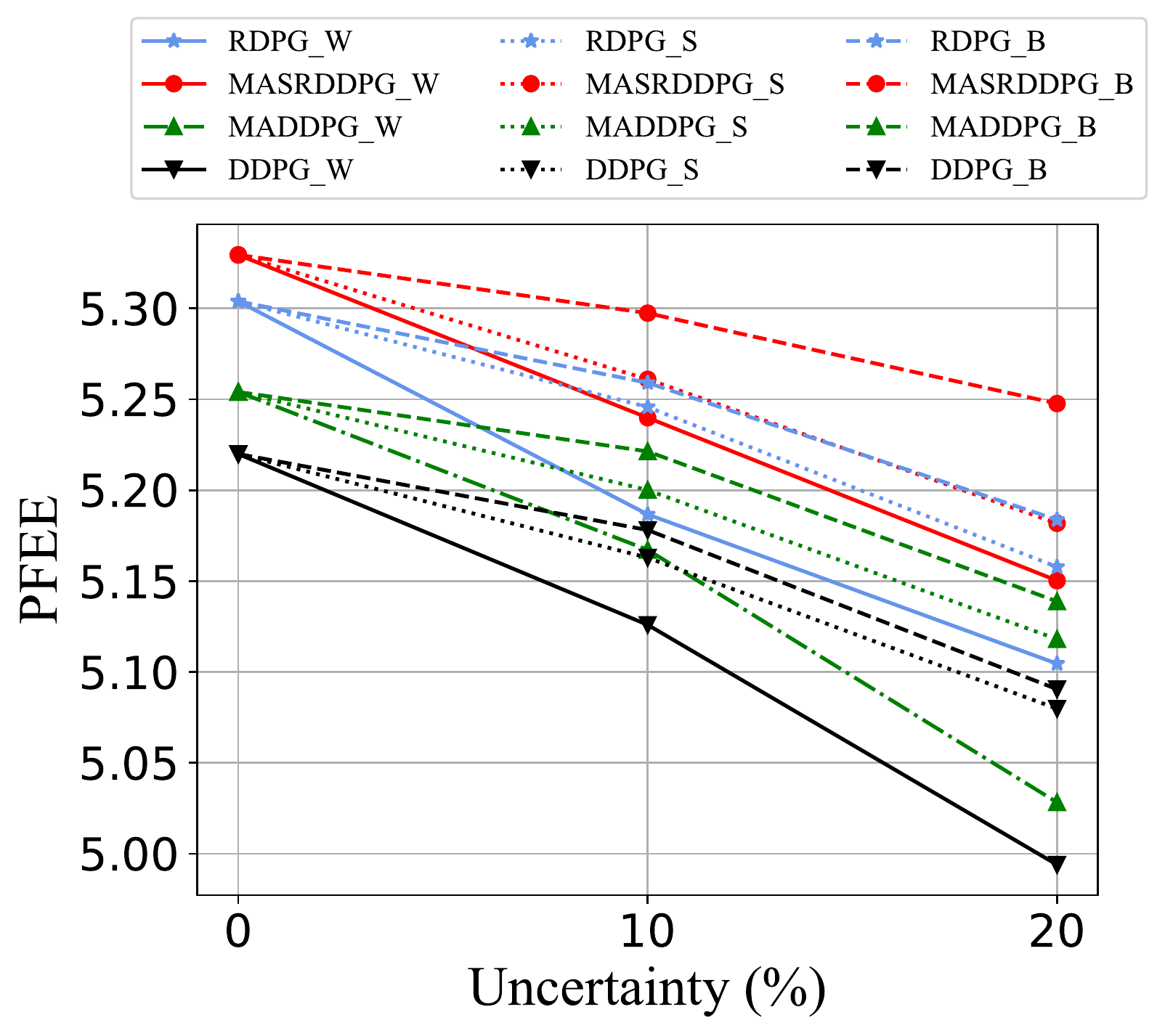}}
		\subfigure[]{\label{sec_sd}\includegraphics[width=5.5cm]{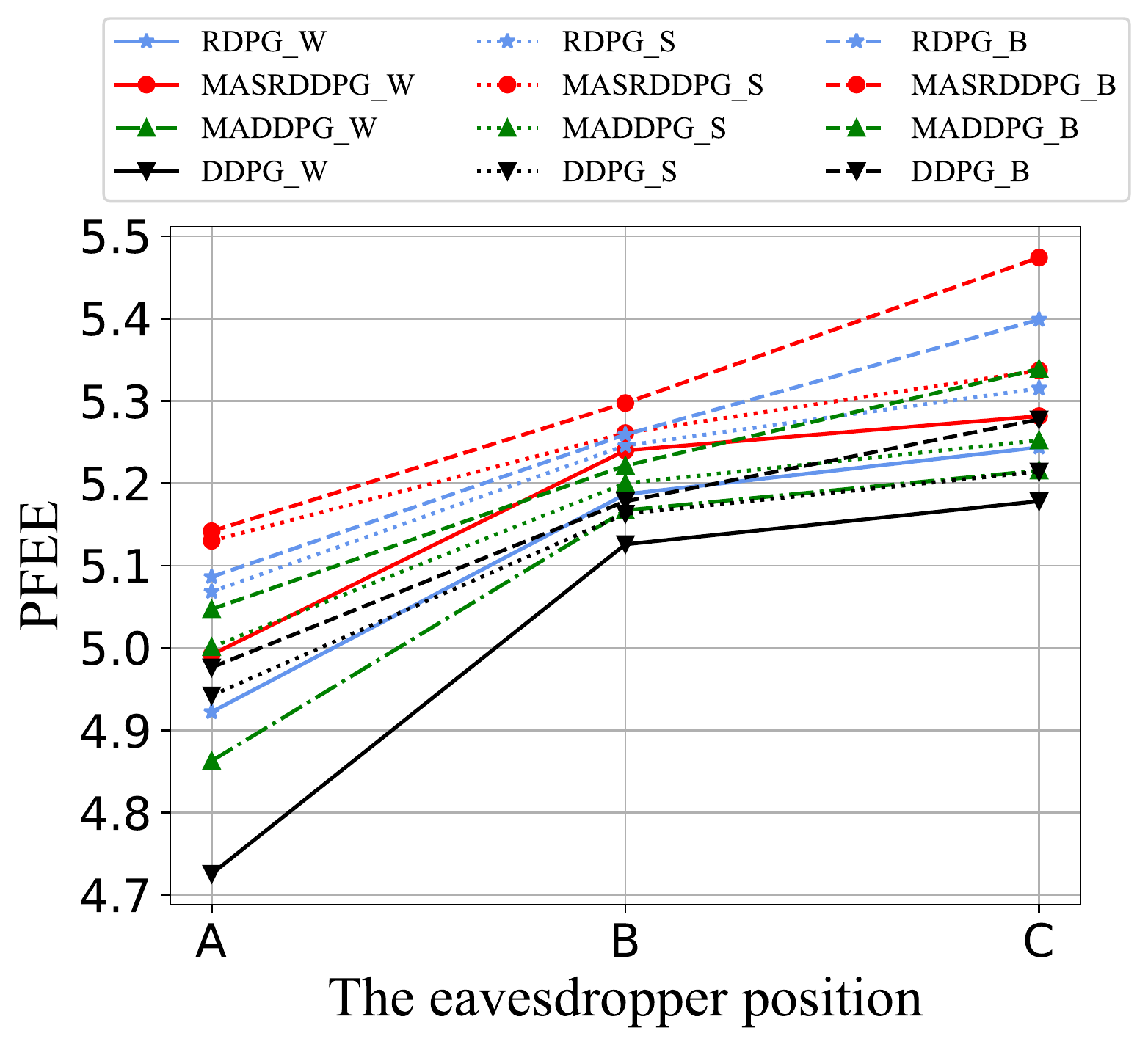}}
		\subfigure[]{\label{sec_bd}\includegraphics[width=5.5cm]{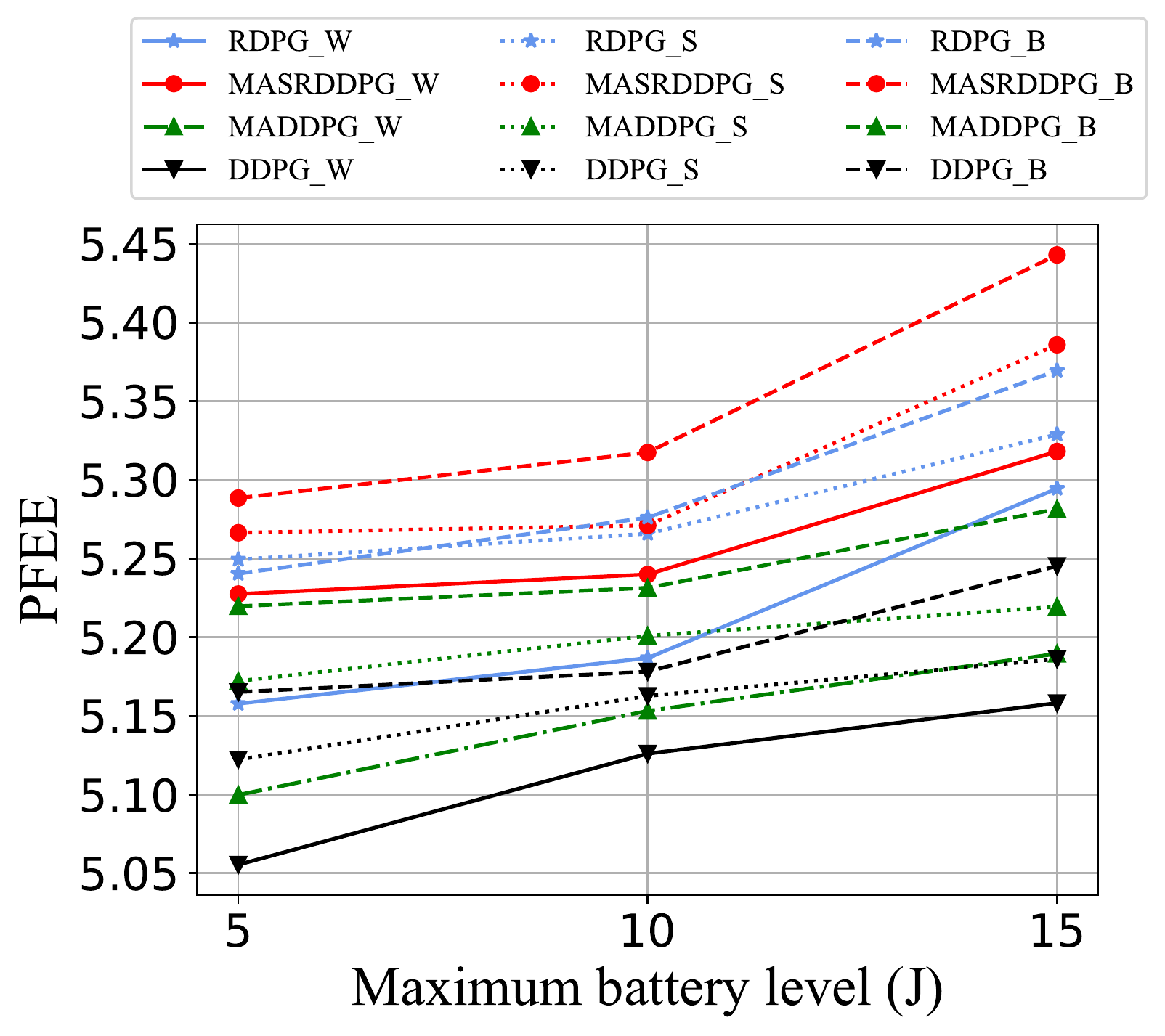}}
		\caption{The PFEE for different methods with worst case "\_W", stochastic "\_S", and Bernstein "\_B" uncertainties. (a) depicts the PFEE versus uncertainty value, (b) depicts versus the eavesdropper position, and (c) depicts versus maximum battery level.} \label{PFEE_total}
	\end{center}
\end{figure}

{	\begin{figure}
		\begin{center} 
			\subfigure[]{\label{sec_wd}\includegraphics[width=5.5cm]{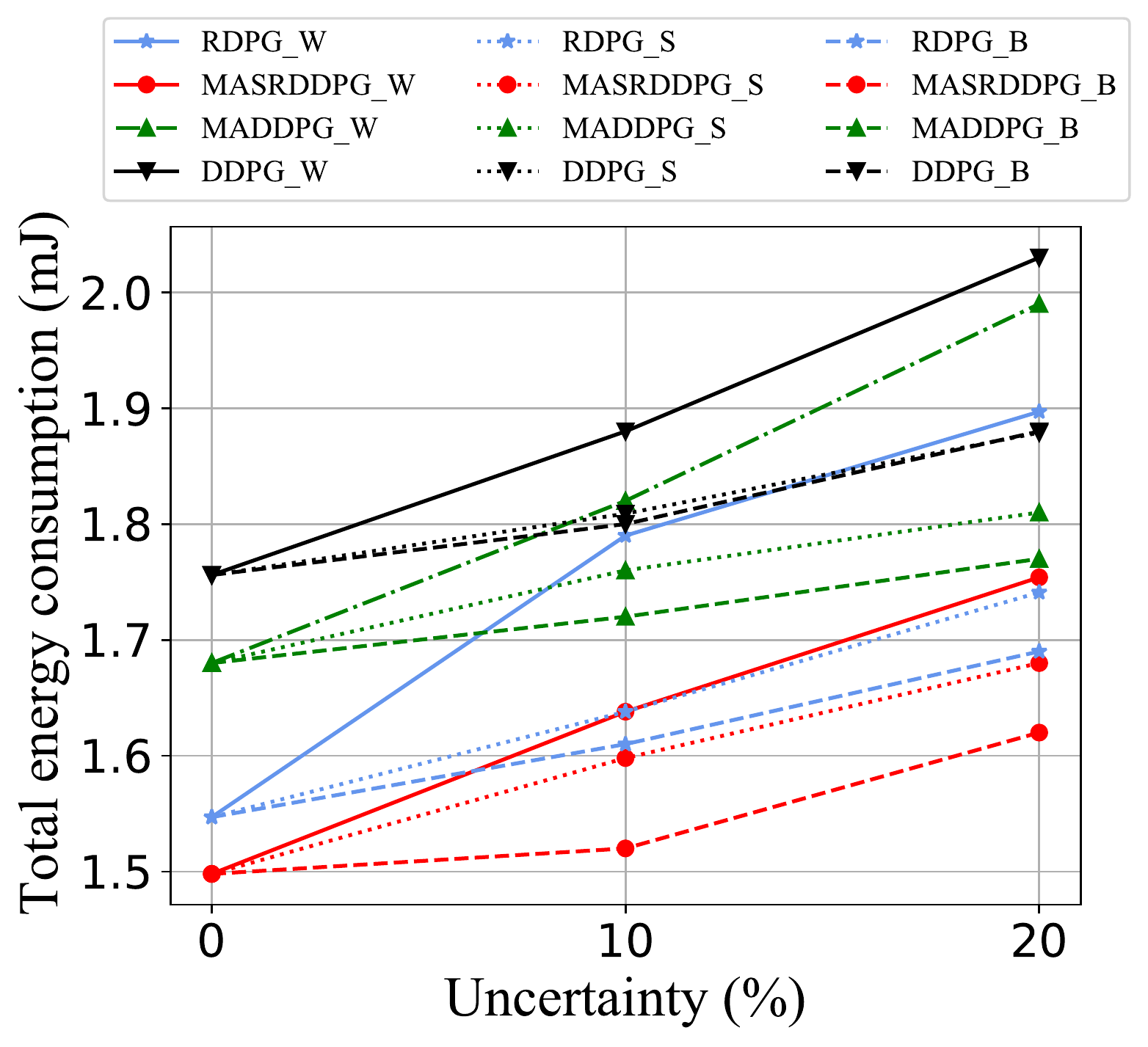}}
			\subfigure[]{\label{sec_sd}\includegraphics[width=5.5cm]{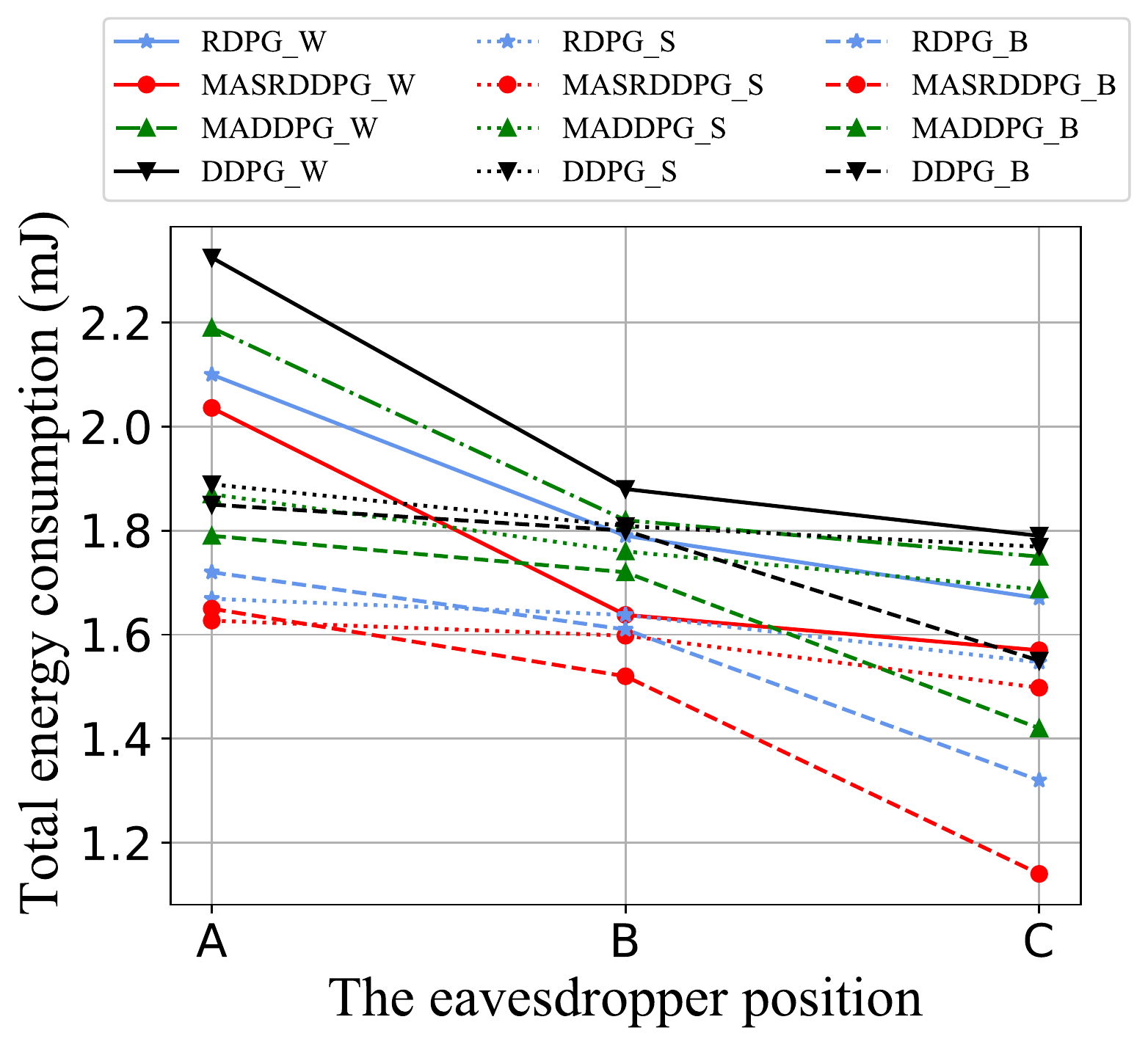}}
			\subfigure[]{\label{sec_bd}\includegraphics[width=5.5cm]{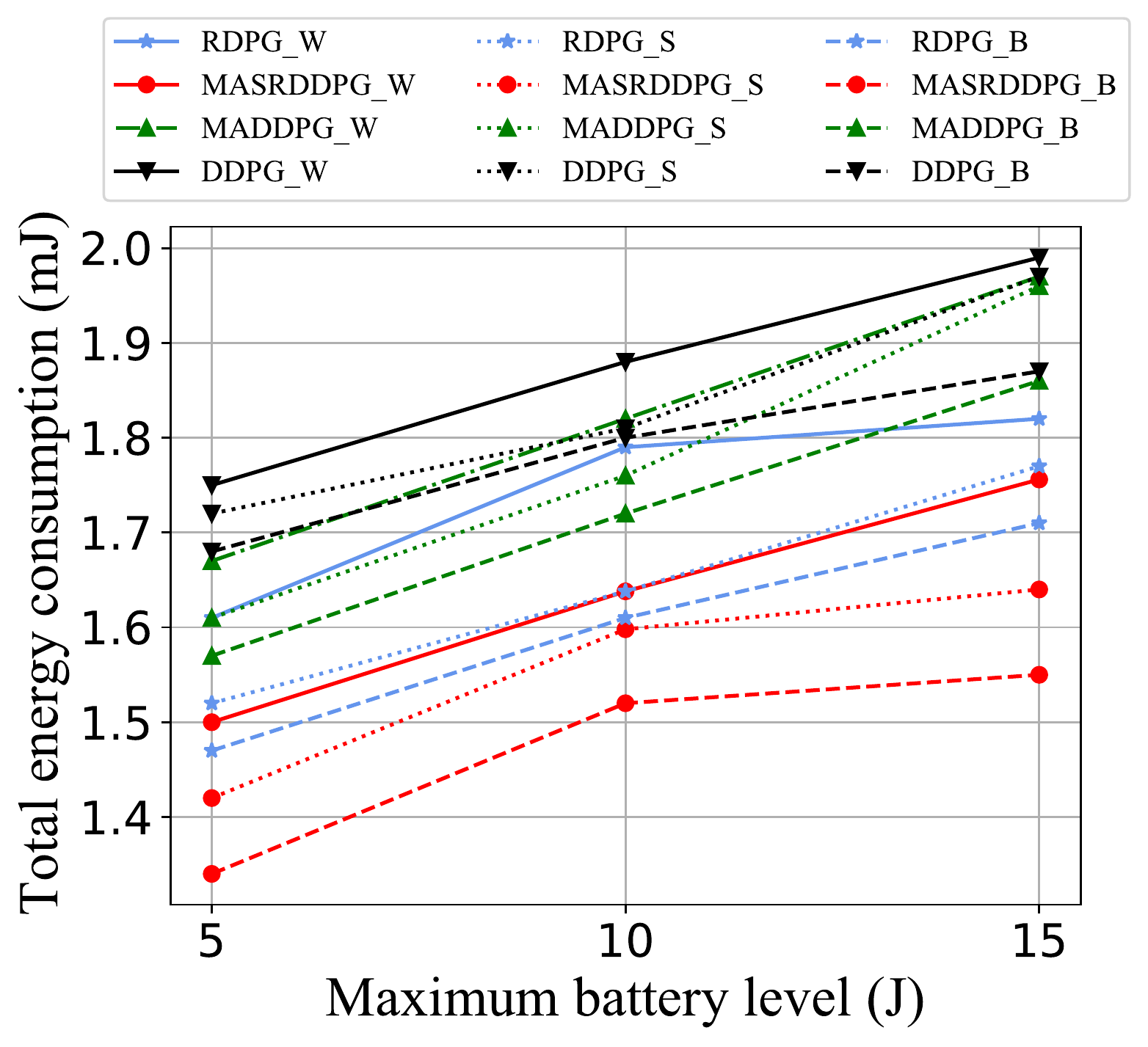}}
			\caption{{The energy consumption for different methods with worst case "\_W", stochastic "\_S", and Bernstein "\_B" uncertainties. (a) depicts the energy consumption versus uncertainty value, (b) depicts versus the eavesdropper position, and (c) depicts versus maximum battery level}.} \label{energy_total}
		\end{center}
	\end{figure}}
	The achieved PFEE for all DRL methods in all three uncertainty models are demonstrated in Fig.~\ref{PFEE_total}. As depicted, regardless to uncertainty model, the MASRDDPG outperforms other DRL methods. The comparison of the average secrecy rate versus energy consumption for all uncertainty models are depicted in Fig.~\ref{SE_EN}. The worst case uncertainty model yields the worst results compared to the stochastic and the Bernstein approximation uncertainty models. In other words, in the worst case uncertainty model, all methods consumed more energy and achieved less average secrecy rate In addition, with the Bernstein uncertainty model, all methods have better performance from energy consumption and the average secrecy rate point of view. The summary of the simulation results are provided in Table \ref{performance}. 

{	\begin{figure}
	\begin{center} 
		\subfigure[]{\label{sec_wd}\includegraphics[width=5.5cm]{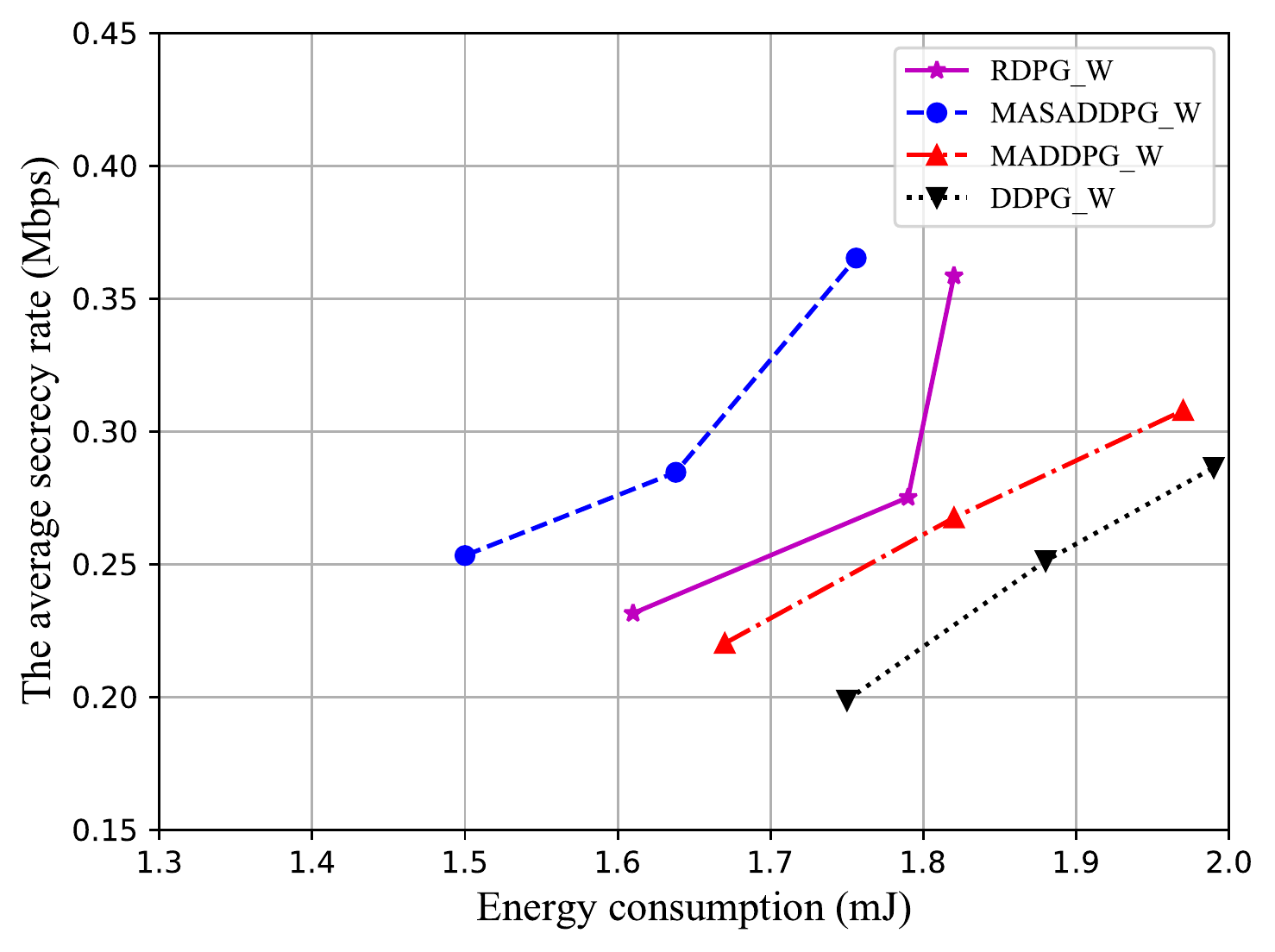}}
		\subfigure[]{\label{sec_sd}\includegraphics[width=5.5cm]{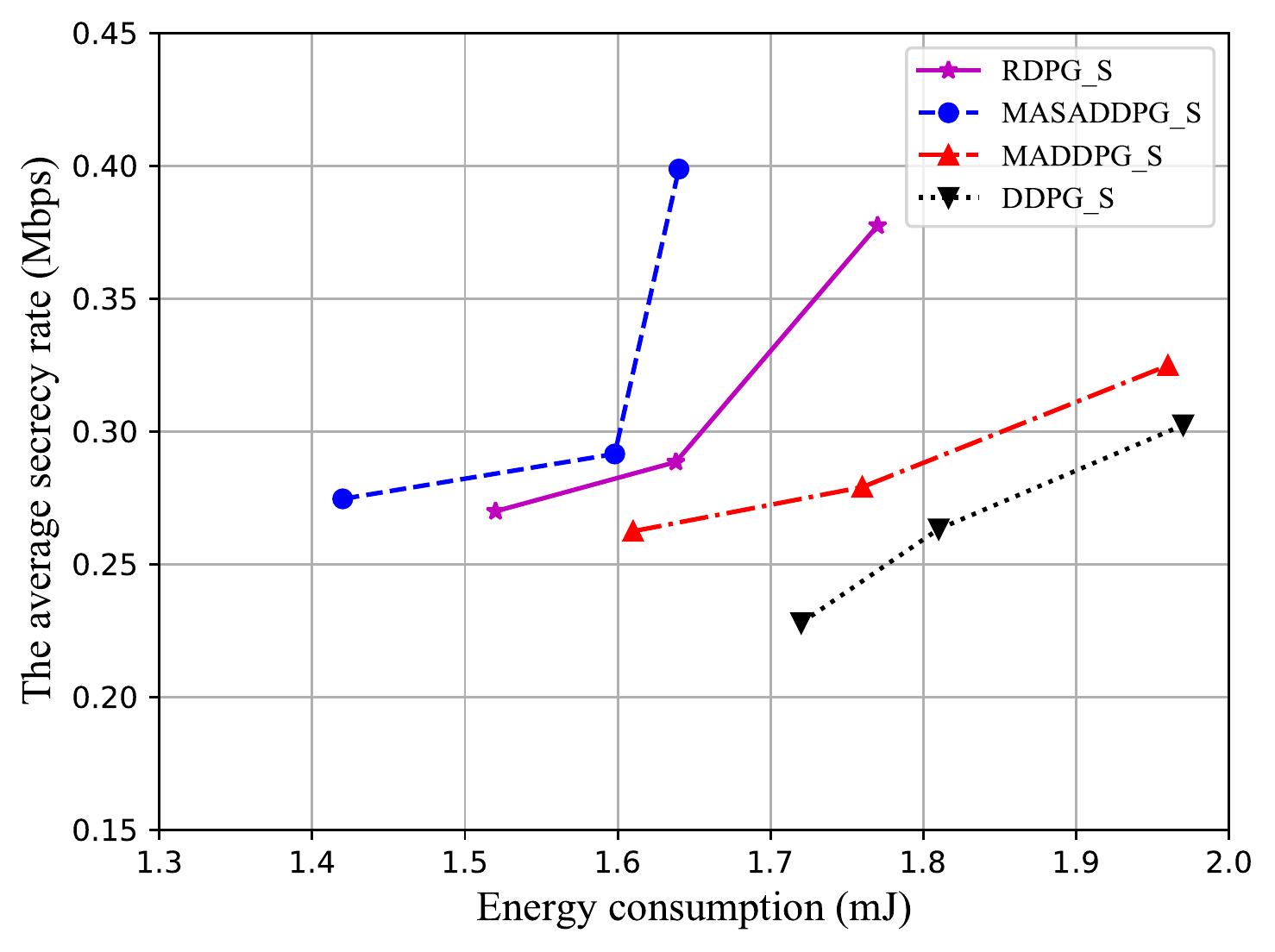}}
		\subfigure[]{\label{sec_bd}\includegraphics[width=5.5cm]{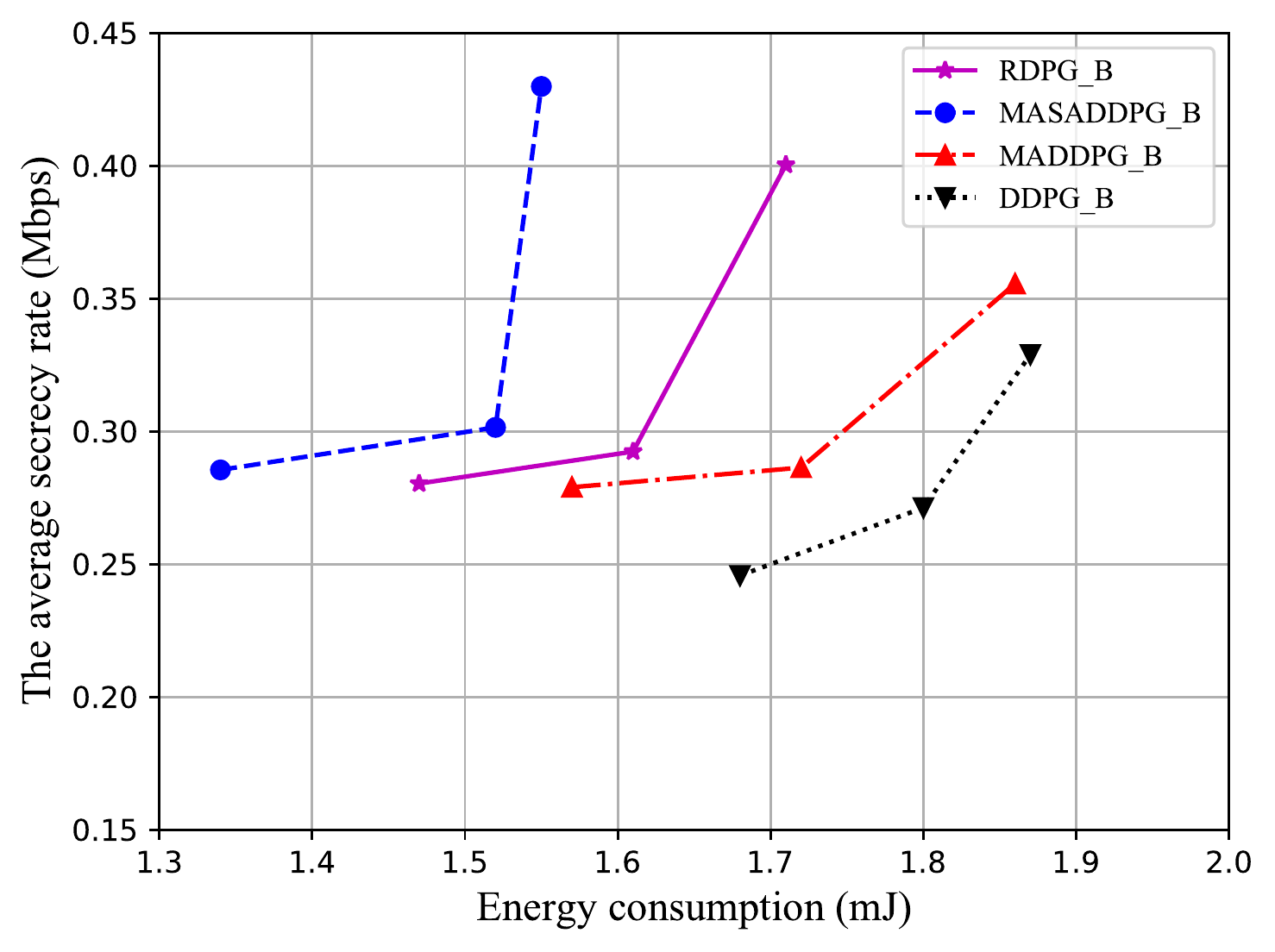}}
		\caption{{The average secrecy rate versus energy consumption with (a) Worst case uncertainty, (b) Stochastic uncertainty, and (c) Bernstein approximation uncertainty.}} \label{SE_EN}
	\end{center}
\end{figure}}

{\begin{table}[tp]
	\
	\caption{{Performance Comparison From The Average Secrecy Rate Point of View}}\label{performance} \centering
	\small
	\begin{tabular}{| p{2.5cm} | p{2.5cm} | p{2.5cm} | p{2.5cm} | p{2.5cm}  |}
		\hline
		\textbf{Method} & Compared to DDPG & Compared to MADDPG &Compared to MASRDDPG & Compared to RDPG \\
		\hline
		\textbf{MASRDDPG} & 31.8\% better & 22.7\% better & -- & 5.6\% better \\
		\hline
		\textbf{RDPG} & 25.2\% better & 16.4\% better & 5.1\% worse & -- \\
		\hline
		
	\end{tabular}%\vspace{-.5cm}
\end{table}

		\section{Conclusion}\label{conclusion}
		As one of the major issues in the energy harvesting cognitive radio, the need for energy efficient secure transmission has gathered more at{tention. We} { pr}oposed a robust and secure resource allocation in cognitive radio based on time switching energy harvesting and PD-NOMA technique, in which two access points as the PAP and the SAP cooperate with each other to serve the PUE with the existence of an malicious user that eavesdrops the data transmitted by the PAP and the SAP. By considering uncertainties for channel gains and battery levels, we formulated our problem of maximizing the PFEE. Then we applied two DRL methods as the MASRDDPG and the RDPG, which are capable to cope with the environments that are uncertain and partially observable. By conducting comprehensive simulations, we demonstrated that the MASRDDPG outperforms RDPG and other state-of-the-arts DRL methods as the uncertainty increases in the network.

		\bibliographystyle{IEEEtran}
		\bibliography{Bibliography}
		\clearpage

	\end{document}